\documentclass[draft]{agujournal2019}
\usepackage{url} 
\usepackage[inline]{trackchanges} 
\usepackage{soul}
\usepackage{mathrsfs}

\linenumbers

\usepackage{bm}
\newcommand{\iu}{{i\mkern1mu}}
\newcommand{\lambdabar}{{\mkern0.75mu\mathchar '26\mkern -9.75mu\lambda}}
\usepackage{amsmath}
\usepackage{amssymb}

\draftfalse

\nolinenumbers

\begin{document}

\title{Modeling lunar response to gravitational waves using normal-mode approach and tidal forcing}

\authors{Josipa Majstorovi\'c\affil{1}, L\'eon Vidal\affil{2}, Philippe Lognonn\'e\affil{1}}

\affiliation{1}{Universit\'e Paris Cit\'e, Institut de physique du globe de Paris, CNRS,
Paris, France}

\affiliation{2}{Universit\'e Paris Cit\'e, Astroparticule et Cosmologie, CNRS,
Paris, France}

\correspondingauthor{Josipa Majstorovi\'c}{josipa.majstorovic@protonmail.com}

\begin{abstract}
In the light of the recent advances in lunar space missions a great interest into using Moon as a future environment for gravitational waves (GWs) detectors has been initiated. Moon offers a unique environment for such detectors due to constrained noise sources, since unlike Earth it does not have ocean and atmosphere. In this paper, we further explore the idea of using Moon as a giant resonator of GWs, a proposal that was first introduced by Weber in 1969. This idea is relaying on the theory how GWs interact with free masses and finally elastic solids, such as is a planet to some approximation. We start by carefully setting up General Relativity (physics) and elastic theory (geophysics) background to be able to derive analytically the coupling between GWs and elastic solids through associated equations of motion. Once the analytical solution is derived, we explore the parameter space this interaction depends on. This eventually provides us with the transfer function, which defines the frequency band of the interest. We show how this interaction robustly depends on the regolith structure by altering the initial lunar model and exploring different regolith models. Our results show that detection might be troublesome in the high frequency regime between 0.1 and 1 Hz, without beforehand constraining the regolith structure with geophysical methods. Finally, we discuss what are the implications of detecting these signals with the future GW detectors build on the Moon. 
\end{abstract}

\section{Introduction}
Nine years after a first detection of gravitational waves (GWs) in 2015 \cite{Abbott2016}, the groundbased detectors, LIGO and VIRGO, created numerous possibility of studying Universe by providing community with catalog of detections \cite{abbott2023gwtc}. Up to this day, LIGO and VIRGO remain the only operating detectors of GWs. It is a well know fact that the sensitivities of these GW detectors are limited by a multitude of noise sources from the instruments themselves and their surroundings \cite{saulson2017fundamentals}. Thus, their exploitable frequency band is defined between $10^1$ and $10^4$ Hz. Although these instruments  are decoupled by several orders of magnitude from the Earth's seismic activity thanks to state-of-the-art isolation systems \cite{SA}, seismic noise prevents any detection below a few Hertz. Furthermore, it is not possible to isolate these detectors from Newtonian noise, which is largely due to our planet's seismic noise and the associated mass re-distribution and results in residual acceleration of the interferometer test masses, degrading performance below 8 Hz. Going in space with the LISA mission, scheduled for the early 2030s, will enable us to explore a lower frequency range from $10^{-4}$ to $10^{-1}$ Hz \cite{LISA_prop}. This mission consists of a constellation of three satellites that will follow the Earth along its heliocentric orbit at a distance of $50 \times 10^6$ km. These satellites shall maintain laser links between them to create three massive interferometers. Additional space-based concepts still under study have been suggested to further investigate frequencies in the milli- to micro-Hertz range \cite{Harry_2006, kawamura2020current, Lacour_2019}. 

During the Apollo missions, seismometers were deployed at the lunar surface which allowed transmitting data back to Earth about the Moon seismic activity from 1969 until September 1977. The analysis of the Apollo data showed that the Moon is extremely seismically quiet with an upper limit on the seismic background noise that is 3 order of magnitude lower than Earth one \cite{https://doi.org/10.1029/2008JE003294}. For this reason, the Moon was and it is still considered as a unique environment for a gravitational astronomy. In addition, GWs are known to excite vibrations of any elastic body and many efforts has been made to study vibrations of various bodies, such as Earth \cite{Weber1959, Dyson1969, Sabbata1970, Ben1983, majstorovic2019earth}, spheres \cite{ashby1975gravitational, Linet1984}, elastic solids \cite{Dozmorov1976a, Dozmorov1976b}, stars \cite{Burke1976, Walgate1983, Khosroshahi1997, Siegel2010, Siegel2011, McKernan2014, Lopes2015, Lopes2017a, Lopes2017b} and recently Moon \cite{kachelriess2023lunar, li2023detecting, yan2024toward, bi2024response}. Subsequently, Weber was the first one to suggest to employ Moon as a giant resonance oscillator of GWs \cite{Weber1959}. Following his idea a gravimeter was deployed at the lunar surface during Apollo 17 \cite{douglass1971new, Giganti1973, Giganti1977}. Unfortunately, due to a technical failure the data could not be used for the search of GWs. Recently, new interest in lunar exploration has led to novel project proposals to detect GWs from lunar surface in order to access the missing deci-Hertz frequency band \cite{harms2022seismic, branchesi2023lunar, cozzumbo2023opportunities}. The Lunar Gravitational Wave Antenna (LGWA) proposal consists of an array of four highend seismometers that shall monitor the Moon's response in the frequency range between $10^{-3}$ to $1$ Hz \cite{harms2021lunar,van_Heijningen_2023}. The Lunar Seismic and Gravitational Antenna (LSGA) proposal involves deploying an antenna configuration composed of two 10 km-long Engineered Fiber optic Distributed Acoustic Sensors (EFDAS) fiber cables arranged in an L-shape, in parallel with two laser strainmeters. The EFDAS cables are interrogated by a central interrogator unit, which utilizes a narrow band laser light source \cite{Katsanevas}. Gravitational-wave Lunar Observatory for Cosmology (GLOC) proposal discusses positioning three end-stations at the rim of a big lunar crater to create a triangle-shape detector (3 independent interferometers) like Einstein Telescope (ET) with arms of 40 kilometers \cite{Jani_2021}. Laser Interferometer On the Moon (LION) proposal consists of a cryogenic triangular interferometer placed in a permanent shadowed region (PSR). Its is based on CE 2 proposal and technologies \cite{Amaro-Seoane_2021, 2021NatRP...3..344B}. 

In the light of the new proposals that consider to establish GW detectors on the Moon surface, it is required to develop a Moon's response to different sources of excitations, as well as Moon response to GWs. The frequency dependent response model to GWs could be used twofold. Either it can be considered as a noise signal that would be removed from the instrument's output to increase instrument sensitivity. Or it could be considered as a signal that we want to measure. Also, depending on what type of measurement we are interested in, the Moon's response should be calculated in carefully chosen coordinate system \cite{harms2022seismic}. Classically, the models of elastic bodies, such as the resonant bar and sphere detectors, are calculated in the proper detector frame, where the GW forcing is considered to be of tidal nature. However, the response of any elastic body can also be derived in the transverse traceless (TT) gauge, where the GW metric forcing is associated with the gradient of the shear modulus. The two approaches derive displacement variables that should be measured with different instruments \cite{ashby1975gravitational}. The motivation of this paper is to focus on the tidal response of the Moon due to GWs, since this approach should be more general and applicable for different type of instruments. Thus, our paper goes into details of tidal response of the realistic lunar model, and completes some of the recently published papers on the same topic \cite{harms2021lunar, branchesi2023lunar, li2023detecting, kachelriess2023lunar, cozzumbo2023opportunities, yan2024toward, bi2024response, ajith2024lunar}.

This paper is organised as follows: first, we introduce the GW metric in TT gauges and the proper detector frame. Second, we discuss how the equations of motion of an elastic body are derived in these two gauges. Third, we show how displacements are derived from these equations for a spherically symmetric, non-rotating Moon in the proper detector frame using the Green tensor formalism. Fourth, we provide analytical radial and horizontal transfer functions. Fifth, we discuss the implications of altering the lunar models on the associated spectra. Finally, we consider the detectability of GWs in the context of the derived analytical solution, instrument sensitivity, and the lunar environment.  

\section{Gravitational wave metric}
In this chapter, we discuss the dynamics of test masses in presence of GW in two different reference frames: Transverse-Traceless (TT) frame and the proper detector frame. The latter will be approximated to a Local Lorentz (LL) frame at the center of mass of our elastic body.
While the physics remains invariant under coordinate transformations, the choice of reference frame does impact the intermediate steps of computations and leads to different interpretation of equations of motion.

\subsection{Transverse-Traceless frame}\label{TT_frame_ch}
Starting from a weak perturbation $h_{\mu \nu}$ of the flat space-time metric $\eta_{\mu \nu}$ with signature (-+++), one can write, in a linearized theory of General Relativity, the metric in a TT gauge as
\begin{equation}\label{eq:dscarre}
    d s^2 =\left(\eta_{\mu \nu}+h^{TT}_{\mu \nu}\right) d x^\mu d x^\nu
\end{equation}
with $h^{TT}_{\mu \nu}$ as a symmetric matrix ($h^{TT}_{\mu \nu} = h^{TT}_{\nu \mu}$). Here and in what follow, we will use Einstein notation, which imply implicit summation as soon as the same index appears in two covariant and contravariant components. Also, Greek letter index varies between 0 and 3 (i.e. space and time), while Latin letter index varies between 1 to 3 (i.e. only in space). Next, the perturbation matrix is transverse ($\partial^j h^{  \; TT}_{i j}=0$), traceless ($h^{i \; TT}_{i}=0$) and does not have time component ($h^{0 \mu \; TT}=0$).
\noindent
Assuming that GWs are propagating along the three-direction produced by far sources we can approximate them as plane waves and therefore describe them with function $x^3-x^0$, where $x^0 = (c* t)$, $c$ being speed of light. This leads to 
\begin{equation}
\begin{aligned}
\nonumber
    h^{TT}_{a b} & =  h_{a b}\left(x^3-x^0\right), \\
    \partial^3 h^{TT}_{a 3} & = 0,
\end{aligned}
\end{equation}
 with $(a, b)=(1,2)$. One can reformulate equation (\ref{eq:dscarre}) and show that the non-vanishing spatial components of the metric in TT are
\begin{equation}\label{eq:ds_TT}
    d s^2=-\left(d x^0\right)^2+\left(d x^3\right)^2+\left(\eta_{a b}+h^{TT}_{a b}\right) d x^a d x^b.
\end{equation}
Before getting an expression of this metric in the proper detector frame using a gauge transformation, it seems important to remind the physical interpretation of the TT frame. 
In order to highligh the effect of GW it is common to describe test masses dynamics using the geodesic deviation equation 
\begin{equation}\label{eq:geod_dev}
    \frac{d^2 \zeta^\mu}{d \tau^2}+2 \Gamma_{\nu \rho}^\mu(x) \frac{d x^\nu}{d \tau} \frac{d \zeta^\rho}{d \tau}+\zeta^\sigma \partial_\sigma \Gamma_{\nu \rho}^\mu(x) \frac{d x^\nu}{d \tau} \frac{d x^\rho}{d \tau}=0
\end{equation}
where $\Gamma_{\nu \rho}^\mu$ is the Christoffel symbol and is restricted to first order in $h_{\mu \nu}$. In the specific case where the separation $\zeta^{\mu}$ between two test masses coordinates $x^\mu$ is small compared to the reduced GW wavelength ($\zeta / \lambdabar \ll 1$), this geodesic deviation equation (\ref{eq:geod_dev}) is valid at first order in $\zeta$.
 Initially, if we assume two test masses at rest in space $d x^i / d \tau=0$, all the movement is in time dimension $d x^0 / d \tau=c$. Therefore, one can compute the coordinate distance $\zeta^{i}$ evolution using the geodesic deviation equation (\ref{eq:geod_dev}) and the expression of Christoffels (see Appendix \ref{eq:christoffel})  in TT gauge becomes
\begin{equation}\label{eq:TT_geo}
    \left.\frac{d^2 \zeta^i}{d \tau^2}\right|_{\tau=0}=-\left[\dot{h}^{TT}_{i j} \frac{d \zeta^i}{d \tau}\right]_{\tau=0}
\end{equation}
This final equation illustrates that if the coordinate speed $\frac{d \zeta^i}{d \tau}$ is equal to zero, it implies that the coordinate acceleration $\frac{d^2 \zeta^i}{d \tau^2}$ is also zero. Consequently, in the TT gauge, the separation $\zeta^{i}$ between two test masses initially at rest will not change over time when a GW passes through: the TT frame will have coordinates marked by test masses. However, as the space-time interval $\mathbf{ds}$ still depend on the GW terms in equation (\ref{eq:dscarre}), one could see the effect of a passing GW through a proper distance measurement, for example by measuring the round trip time taken by a light beam between two test masses. As discussed in \citeA{Maggiore2008}, the proper distance $\mathbf{s}$ in TT can be written at first order in $\mathrm{h}$ as 
\begin{equation}
    s \simeq L+h^{TT}_{i j}\left(L_i L_j / 2 L\right),
\end{equation}
with $\mathbf{L}$ the coordinate distance between these two test masses.

\subsection{Proper detector frame}\label{LLframe}
The GW metric can also be expressed in other coordinate systems, which will result in a different interpretation of test masses motions in presence of GW. Following the method used by \citeA{ashby1975gravitational}, we perform a gauge transformation (see Appendix \ref{annexe:TT_to_LL}) to establish a local inertial frame at Moon's center of mass. Assuming that the center of mass follows a geodesic of space-time, a convenient way to describe our problem is to put the origin of our reference frame there and describe the problem using Fermi normal coordinates \cite{Maggiore2008}. Using the latter, the metric tensor at the origin of the Fermi coordinates takes the form of the Minkowski metric and the metric tensor can be expanded as a Taylor series around this particular worldline. One can show that  the second-order terms of this expansion makes the Riemann curvature tensor appear \cite{Maggiore2008}. This allows us to see the effects of curvature in the local region around the origin of our frame. Another interest to this is that, as we will see in the following, in Fermi coordinate the GW field is naturally represented as a Newtonian force. The transformation from TT to Fermi coordinate leads a to new expression of the  metric as 
\begin{equation}\label{eq:ds_LL}
    \begin{aligned}
d s^2= & -\left(d \bar{x}^0\right)^2\left(1-\frac{1}{2} h_{a b, 00} x^a x^b\right)+\left(d \bar{x}^3\right)^2\left(1+\frac{1}{2} h_{a b, 33} x^a x^b\right) \\
& +2\left(\frac{1}{2} h_{a b, 03} x^a x^b d \bar{x}^0 d \bar{x}^3\right)+\left(d \bar{x}^1\right)^2+\left(d \bar{x}^2\right)^2
    \end{aligned}
\end{equation}
where we omit the TT index on $h$ for more clarity. This formulation highlights the fact that the metric depends quadratically on the distance $x^a$ with respect to the center of our frame. Consistent with the properties of Fermi coordinates, we obtain on the submanifold $x^a = 0$ a flat metric ($g_{\mu \nu}=\eta_{\mu \nu}$) which can be associated to a local Lorentz (LL) frame where $\Gamma_{ \nu \rho}^\mu=0$ is valid.
\\
As we did previously in TT frame, one can use equation (\ref{eq:geod_dev}) to interpret this new metric in the proper detector frame. In the case of a non-relativistic object like the Moon, we can neglect the spatial four-velocity components with respect to its time component \\ ($d x^i / d x^0  \ll 1$). Therefore, the third term of the geodesic deviation equation will only depend on $\partial_\sigma \Gamma_{0 0}^\mu(x)$ and, since the metric components does not depend on $x^0$ or $x^3$, only the transverse spatial derivatives will remain. The second term, depending on $\Gamma_{ \nu \rho}^\mu$, also vanishes in equation (\ref{eq:geod_dev}). The evolution of spatial coordinate distance between two test masses is then obtained with expression
\begin{equation}\label{eq:geo_LL}
    \frac{d^2 \zeta^i}{d \tau^2}+\zeta^j \partial_j \Gamma_{00}^i\left(\frac{d x^0}{d \tau}\right)^2=0.
\end{equation}
It is common in the literature to express the second term of equation (\ref{eq:geo_LL}) with the Riemann curvature tensor. By identification and using the fact that the time derivative of the Christoffel symbols are zero here ($\partial_0 \Gamma_{0 j}^i=0$ ), one can recognize at first-order in $h$ as
\begin{equation}
R^i_{ 0 j 0} =\partial_j \Gamma_{00}^i  = -\frac{1}{2} h_{a b, 00} =- \frac{1}{2c^2} \ddot{h}_{a b}.
\end{equation}
At first order in $h$, we can use $\frac{d x^0}{d \tau} = c$ and express equation (\ref{eq:geo_LL}) with the double time derivative of $h_{ab}$. Because the $z$-axis was
previously defined as the propagation axis in Section 2.1, the non vanishing terms of the Riemann tensor depends only on the transverse components of $h$ and the relative movement of the test masses will be describe in the orthogonal plane to $z$ as
\begin{equation}
    \frac{d^2 \zeta^a}{d \tau^2} = \frac{1}{2}\zeta^b \ddot{h}_{a b},
\end{equation}
\noindent
which signifies a tidal acceleration induced by propagating GW in the proper detector frame. This result agrees well with \citeA{ashby1975gravitational}, however with different sign.\\

Therefore, the influence of GWs in the proper detector frame can be interpreted as a "Newtonian" tidal force $F_a$ acting between test masses of mass $m$, depending on the double time derivative of metric $\ddot{h}$ as
\begin{equation}\label{eq:tidal_force}
    F_a=\frac{m}{2}\zeta^b \ddot{h}_{a b}. 
\end{equation}

\section{Equations of motion}
After defining how gravitational waves interact with test masses, we need to consider their effects on an elastic medium and include the elasto-dynamic counterparts in the previously defined geodesic equations. \citeA{Dyson1969} used a non-relativistic field approach for a flat Earth model, followed by \citeA{Ben1983} for a spherical Earth, in order to get an equation of motion in TT for an elastic medium coupled with GW. In  both papers, the gravitational self-field of the planet is neglected. In an elastic medium, it is common to describe the displacement of each part of the medium from its equilibrium position $x$ with a displacement field: we distinguish its notation between TT and detector frame, choosing respectively $\bm{\xi}(t, \mathbf{x})$ and $\mathbf{u}(t, \mathbf{x})$, respectively. In the following, we will omit the time and spatial dependency in the notation for more clarity. The elasto-dynamic equation when no external sources are active, nor gravity is accounted for, can be written as
\begin{equation}
    \label{eqn:vol}
    \frac{\partial}{\partial t} \left(\rho \dot{\xi}_i\right) -\nabla^j \sigma_{i j}=0,
\end{equation}
where $\sigma_{ij}$ is the elastic stress tensor, $\rho$ the density, $\nabla^j$ derivation along j and where the associated boundary conditions are defined as
\begin{equation}
    \label{eqn:bound}
    n^j \sigma_{i j}=0,
\end{equation}
where $n^j$ is the vector normal to the surface. Assuming an isotropic medium, the stress tensor can be written as 
\begin{equation}
    \sigma_{i j} (\bm{\xi})=\lambda(\mathbf{x}) \delta_{i j} \nabla^k \xi_k+\mu(\mathbf{x})\left(\nabla_i \xi_j+\nabla_j \xi_i\right),
\end{equation}
where $\lambda$ is the Lamé constant, $\mu$ is the shear modulus, and $\nabla^k \xi_k = \partial^k \xi_k$.

\citeA{Dyson1969} used a Lagrangian approach to express the perturbation generated by GW and have shown that both the equation of motion ($\ref{eqn:vol}$) and the boundary condition ($\ref{eqn:bound}$) are modified with the addition of a volume and surface density force, respectively. After linearisation and to the first order, this leads to the expressions as follows
\begin{eqnarray}
    \label{eq:motion_TT_non_relat}
    & \rho_0  \ \ddot{\xi}_i  = \nabla^j \sigma_{i j} (\mathbf{\xi}) + f_i^{\mathrm{Dyson}}, \\
     & n^j \left( \sigma_{i j}  +  t_{i j}^{\mathrm{Dyson}} \right) = 0 ,
\end{eqnarray}
where $\rho_0$ is the equilibrium density and with the GW excitation forces defined as
\begin{eqnarray}\label{eq:force_dyson}
  & f^{\mathrm{Dyson}}_i =  \nabla^j ( \mu  \ h_{i j}^{\mathrm{TT}} ) , \\
  & t_{i j}^{\mathrm{Dyson}} =   \mu \  h_{i j}^{\mathrm{TT}} .
\end{eqnarray}
In this description, we see that the GW forcing term emerges from the divergence of the medium's shear modulus in the equation of motion. This implies that GW couples through $h_{i j}^{\mathrm{TT}}$ only with internal discontinuities and external boundary within a solid body. Sign difference compared to \citeA{Dyson1969} derives from the different definition of the metric in terms of the covariant and contra-variant notation \cite{ashby1975gravitational, belgacem2024coupling}. As recently noted in \citeA{belgacem2024coupling}, Dyson's description is valid only when non-relativistic mechanics is used to describe the motion of the solid. Note that for a rigorous description within the framework of the relativistic theory of elasticity, equation (\ref{eq:motion_TT_non_relat}) needs to be revisited.

For a description in the proper detector frame, the force defined with equation (\ref{eq:tidal_force}) couples directly to the density as a Newtonian force \cite{Maggiore2008}, instead of coupling with the shear modulus divergence. An inhomogeneous equation of motion for the associated displacement field $\mathbf{u}$ can be directly obtained as
\begin{equation}\label{eq:EQM_LL}
    \rho_0  \ \ddot{u}_i  = \nabla^j \sigma_{i j} (\mathbf{u}) + f_i^{\mathrm{tidal}} , 
\end{equation}
with a tidal density force term as
\begin{equation}\label{eq:tidal_start}
   {f}^{\mathrm{tidal}}_i =  \frac{1}{2} \rho(\mathbf{x}) \ddot{h}_{i j} x^j. 
\end{equation}
This formulation is only valid in the case of a small body size compared to the GW reduced wavelength, therefore it could be applied to the Moon when searching for GW in the mHz band.

In the case of a self-stressed body like the Moon, press-stress and gravity acceleration $\mathbf{g}$ need to be taken into account. Then, the equation of motion can be written in a general form as 
\begin{equation}\label{eq:tidal_main}
   \rho_0 \ddot{\mathbf{u}} =\mathbf{A}(\mathbf{u}) + \mathbf{f}
\end{equation}
using the linearized gravito-elastic operator $\mathbf{A}$ \cite{lognonne1998computation} in vector form as
\begin{equation}\label{eq:SNREI_1}
     \mathbf{A} (\mathbf{u}) =\frac{1}{\rho_0}[\mathbf{\nabla} \cdot \mathbf{\sigma}(\mathbf{u})+ \rho_1 \mathbf{g_0} + \rho_0 \mathbf{g_1}] , 
\end{equation}
with the associated expressions 
\begin{equation}\label{eq:SNREI_2}
\begin{aligned}
     \rho_1 & = - \mathbf{\nabla} \cdot (\rho_0 \mathbf{u}) ,  \\
     \mathbf{\sigma}(\mathbf{u}) & =  -   \mathbf{u} \cdot   \mathbf{\nabla}  \mathbf{\sigma}(\mathbf{u}_0) + \mathbf{\sigma}_{Hooke}(\mathbf{u}) ,  \\
     \mathbf{\nabla} \cdot  \mathbf{g}_1 & = -4 \pi \cal{G} {\rho}_1 . \\
\end{aligned}
\end{equation}
Next, we will assume that the gravity terms remain small compared to the elasticity ones and that no gravity changes is generated by the GW metric variation. Thus we can use the force term described in the detector frame when the gravito-elastic equation is used. To solve equation (\ref{eq:tidal_main}) with a GW force term, we first solve the free-source equation, generally by computing the eigensolutions, which are known to be a complete basis for all admissible solutions with excitation terms \cite{Dahlen1998,lognonne1998computation}. Complete basis of eigensolution can be used to compute the equation of motion for a point space and time source. This is expressed as a Green tensor. Using this formalisam we can compute any particular solution of equation (\ref{eq:tidal_main}) by contracting the force term with the obtained Green tensor solution. The details will follow in the Section \ref{lunar-section}. 

\subsection{Interpretation and discussion}
It is important to discuss the connections between equations (\ref{eq:motion_TT_non_relat}) and (\ref{eq:EQM_LL}). We can connect $\bm{\xi}$ and $\mathbf{u}$ by analyzing a measurement system in the two coordinate systems defined previously. This will help show that the Dyson equation is directly related to what a seismometer measures. 

An ideal seismometer consists of a free mass $\mathbf{TM_{seis}}$, or inertial mass, attached by a spring to a frame that moves along with any motion of the ground. By measuring the relative displacement between the mass and frame, one can get data on ground vibration. Much above the resonance frequency of the seismometer, the proof mass is, however, decoupled from the ground and is assumed to be in free fall, as it is for VIRGO or LIGO GW detectors.   

For our two coordinate systems (TT and the proper detector frame), we define the origin of our reference frame at the Moon's center of mass and we place at this particular point a reference test mass $\mathbf{TM_0}$. We also place our seismometer at the surface of the Moon at a distance $\mathbf{R}$ from its center.

As this point we can pose a question: in the proper detector frame, what is the relative motion between the seismometer's frame and the inertial mass $\mathbf{TM_{seis}}$ with respect to $\mathbf{TM_0}$? The frame's motion, co-moving with the ground, is described by the displacement field $\mathbf{u^{\textbf{frame}}}$, which is a solution to equation (\ref{eq:EQM_LL}). When the free fall conditions apply, the motion of the inertial mass, independent of the ground's motion, is described by $\mathbf{u^{\textbf{TM/seis}}}$ and is only determined by the gravitational wave field as
\begin{equation}
    \rho(\mathbf{x}) \ddot{u}^{\textbf{TM/seis}}_i=\frac{1}{2} \rho(\mathbf{x}) \ddot{h}_{i j} x^j,
\end{equation}
which correspond to equation (\ref{eq:EQM_LL}) where $\lambda = \mu = 0$. This leads to a displacement given by $u^{TM/seis} = \frac{1}{2}hR$ at the moon's surface. The difference between $\mathbf{u^{\textbf{frame}}}$ and $\mathbf{u^{\textbf{TM/seis}}}$ give us the relative measurement made by our ideal seismometer which is $\mathbf{u^{\textbf{frame}} - \frac{1}{2}hR}$, as acknowledged in \citeA{ashby1975gravitational,yan2024toward, belgacem2024coupling} by writing  
\begin{equation}\label{eq:seis_meas}
\bm{\xi}=\mathbf{u} - \frac{1}{2}h\mathbf{R}.     
\end{equation}
Thus, this displacement is directly related to the response of the Moon to GW. In the TT frame, although the coordinate distance does not change between $\mathbf{TM_{seis}}$ and $\mathbf{TM_0}$, as mentioned in Section \ref{TT_frame_ch}, the proper distance varies when a gravitational wave passes. 

Next, we can ask what would be the the ground's motion relative to $\mathbf{TM_{seis}}$? It's determined by equation (\ref{eq:motion_TT_non_relat}) where the coordinate grid is defined by test mass positions. The variable $\bm{\xi}$ is the one to consider in order to get the local inertial acceleration experienced by our ideal seismometer. In order to pass from equation (\ref{eq:motion_TT_non_relat}) to (\ref{eq:EQM_LL}), one could substitute relation (\ref{eq:seis_meas}) in (\ref{eq:motion_TT_non_relat}), as it was first shown in \cite{ashby1975gravitational}. This follows from the fact that $\mathbf{\nabla} \cdot \mathbf{u}  = \mathbf{\nabla} \cdot \bm{\xi}$, since in the TT gauge $h$ is both transverse and traceless.

To preserve a global description of lunar deformation due to gravitational waves, rather than depending solely on seismometer measurements, we will consider the proper detector frame and the displacement field $\mathbf{u^{\textbf{frame}}}$. This field will also be the most adapted for measuring strain between two points on the lunar surface. We will describe the effect of the gravitational wave as a tidal force (equation (\ref{eq:EQM_LL})) instead of a coupling with the gradient of the shear modulus (equation (\ref{eq:motion_TT_non_relat})) in the remainder of this article.

\section{Lunar response}\label{lunar-section}
In this chapter we develop a lunar response to a GW by considering our system to be in the proper detector frame. This implies solving equation (\ref{eq:tidal_main}). We do not solve the equation directly but use the Green tensor formalism and the normal-mode summation theory following \citeA{Dahlen1998}. 
\subsection{Theoretical background}
The Green tensor, conventionally written as $\mathbf{G(\mathbf{x}, \mathbf{x}'; t)}$, describes the displacement response at position $\{\mathbf{x}$, $t \}$ by an impulsive point-force acting at the position $\{ \mathbf{x}'$, $0 \}$. We can find $\mathbf{G(\mathbf{x}, \mathbf{x}'; t)}$ by solving the equation of motion with the inhomogeneous initial conditions: a) no displacement, b) the velocity is $\frac{1}{\rho_0} \mathbf{I} \delta (\mathbf{x} - \mathbf{x'})$, where $\mathbf{I}$ is the impulse of the point-force and $\delta$ is the 3D Dirac function. Next, we consider the non-rotating, gravito-anelastic operator (\ref{eq:SNREI_1}), and treat anelasticity at the zero-order level in perturbation theory. Thus, we consider only a perturbation to the real frequency by introducing an additional imaginary component. Therefore, the Green tensor is defined as 
\begin{equation}\label{eq:green}
    \mathbf{G} ( \mathbf{r}, \mathbf{r'}; t) = \Re \sum_{k} (\iu \nu_k)^{-1} \mathbf{s}_k (\mathbf{r}) \mathbf{s}_k (\mathbf{r'}) \mathrm{e}^{\iu \nu_k t},
\end{equation}
where $\Re$ is the real part, $k$ stands for $\{n,\ell,m\}$ indices with $n$ being the radial order, $\ell$ the angular order and $m$ the azimuthal order and is defined in the range $\{-\ell:\ell\}$, $\nu_k$ stands for a complex frequency that contains the eigenfrequencies $\omega_k$ and the decay rate $\gamma_k$ as $\nu_k = \omega_k + i \gamma_k = \omega_k + i \frac{\omega_k}{2 Q_k}$, and $\mathbf{s}_k$ stands for the associated eigenfunctions. The associated displacement eigenfunction define a complete orthonormal basis and are defined as
\begin{eqnarray}
    \mathbf{s}_k (\mathbf{r}) = \sum_{\alpha = 0,1,2} \mathcal{D}_{ln}^{\alpha}(r) \mathcal{P}_{lm}^{\alpha}(\theta, \phi)
\end{eqnarray}
which can expand to 
\begin{equation}\label{eq:eigenfunctions}
\mathbf{s}_k (\mathbf{r}) = \mathcal{U}_k(r) \mathbf{\hat{e}}_r \mathcal{Y}_{\ell m} (\theta, \phi) + k^{-1} \mathcal{V}_k(r) \nabla_{1} \mathcal{Y}_{\ell m} (\theta, \phi) - k^{-1} \mathcal{W}_k(r) (\mathbf{\hat{e}}_r \times \nabla_{1} \mathcal{Y}{_\ell m}(\theta, \phi)),
\end{equation}
where $\mathcal{U}_k(r),\mathcal{V}_k(r),\mathcal{W}_k(r)$ are the complex eigenfunction depending on $r$, $\mathcal{Y}_{\ell m}$  the real spherical harmonics (see Appendix \ref{harmonics}), and $k = \sqrt{\ell(\ell+1)}$. As noted above, to the zero-order of perturbation theory anelasticity is retained only upon the eigenfrequencies, thus for the remainder of this paper we use the real eigenfunction $U_k, V_k, W_k$ that depend only on $\ell$ and $m$. The more general cases with attenuation and no rotation or with attenuation and rotation, are described by \citeA{tromp1990free} and \citeA{lognonne1991normal}, respectively. These are however non-necessary complexities due to the weak attenuation of the Moon and its very slow rotation \cite{gillet2017scattering}. We follow the normalisation defined as 
\begin{equation}\label{eq:norm}
\begin{aligned}
    \int_{0}^{R_{P}} \rho  \ ( {}_{n}U_{\ell} \ {}_{n'}U_{\ell} + {}_{n}V_{\ell} \ {}_{n'}V_{\ell} ) \  r^2 dr & = \delta_{nn'} \\
    \int_{0}^{R_{P}} \rho  \  {}_{n}W_{\ell}  \ {}_{n'}W_{\ell} \ r^2 dr & = \delta_{nn'}
\end{aligned}
\end{equation}
with $R_{P}$ being the lunar mean radius and the real spherical harmonics normalization 
\begin{equation}\label{eq:spherical}
    \int_{2 \pi}  d\phi \int_{\pi} \sin \theta d \theta  \ \mathcal{Y}_{\ell m} (\theta, \phi) \mathcal{Y}_{\ell'm'} (\theta, \phi)   = \delta_{\ell\ell'} \delta_{mm'}.
\end{equation}
Usually, we can obtain $U_k(r),V_k(r), W_k(r),\nu_k$ for a specific radially dependent model of a planet, defined by the compressional ($v_p$) and shear ($v_s$) velocities and by the density ($\rho$) radial profiles by a numerical solver such as MINEOS \cite{mineos-v1.0.2}. The displacement of a particle away from its equilibrium position at time $t$ due to an equivalent body force $\mathbf{f}$ and an equivalent surface force $\mathbf{t}$ using the Green tensor formalism can be defined as  
\begin{equation}\label{eq:displacement}
\mathbf{s} (\mathbf{r},t) = \int_{-\infty}^{t} \int_{V} \mathbf{G}( \mathbf{r}, \mathbf{r'}; t - t') \cdot \mathbf{f} (\mathbf{r'}, t') dV' dt' + \int_{-\infty}^{t} \int_{S}  \mathbf{G}( \mathbf{r}, \mathbf{r'}; t - t') \cdot \mathbf{t} (\mathbf{r'}, t') d\Sigma' dt',
\end{equation}
where the first integral is over the volume, while the second one over the surface. To solve equation (\ref{eq:displacement}) we need to define the GW excitation force. Therefore, one needs to understand how the GW is coupled with the elastic medium, that is the solid elastic body as described in previous chapter. We are interested in developing a solution in a proper detector frame as explained in previous chapter, thus we redefine force term defined by equation (\ref{eq:tidal_start}) as
\begin{equation}\label{eq:tidal}
\mathbf{f}_{tidal} (\mathbf{r},t)= \frac{1}{2} \rho (r) r \mathbf{\hat{e}}_r \cdot \mathbf{\ddot{h}(t)}, 
\end{equation}
with $\mathbf{\hat{e}}_r$ being the radial unit vector and $\mathbf{h(t)}$ being the metric perturbation.  We assume that our source is monochromatic, therefore limiting ourselves to a particular GW source, such as are the inspiraling white dwarf binaries. The formula (\ref{eq:tidal}) suggests $\omega_g^2$ dependency of the metric perturbation, where $\omega_g$ is the GW wave frequency, therefore we can introduce 
\begin{equation}\label{eq:GWsource}
    \mathbf{\ddot{h}}(t) = - h_0 \omega_g^2 \bm{\epsilon} \mathrm{e}^{i (\omega_g t - \mathbf{k}_g \cdot \mathbf{r})}, 
\end{equation}
where $h_0$ is the GW strain amplitude, $\bm{\epsilon}$ is a polarization tensor adopted from \citeA{majstorovic2019earth}, $\mathbf{k}_g = \frac{\omega_g}{c} \mathbf{\hat{e}}_k $ is the wave number with $c$ being the speed of light and $\mathbf{\hat{e}}_k$ a unit vector normal to the wave front of the GW, where we set $\mathbf{\hat{e}}_k = \mathbf{\hat{e}}_r$. One needs to keep in mind that in the case of GW source different from the infinitive monochromatic one, one should redefine equation (\ref{eq:GWsource}). Next, with the tidal forcing we associate a boundary condition so that the total force per unit area at the surface vanishes in the direction normal to the surface (see equation (\ref{eqn:bound})), which suggests $ \mathbf{t}=0$ in equation (\ref{eq:displacement}). Thus, we need to compute 
\begin{equation*}
\mathbf{s} (\mathbf{r},t) = \int_{-\infty}^{t} \int_{V} \mathbf{G}( \mathbf{r}, \mathbf{r'}; t - t') \cdot \frac{1}{2} \rho (r') r' \mathbf{\hat{e}}_r \cdot \mathbf{\ddot{h}(t')} dV' dt', 
\end{equation*}
and to obtain the spheroidal displacement, we additionally set $W_k = 0$ in equation (\ref{eq:eigenfunctions}). In the long-wavelength regime ($\omega_g$ $R_{planet}$/c $<<$ 1)  we finally get: 
\begin{equation}\label{eq:disp_tidal}
\mathbf{s}_k (\mathbf{r},t) =  - \frac{1}{2} h_0 \omega_g^2 \mathbf{s}_k (\mathbf{r}) \bar{g}_k(t) \  \bm{\epsilon} :  \left[\int_{V} r' \  \mathbf{\hat{e}}_{r'} \otimes \mathbf{s}_k (\mathbf{r}')  dV' \right],
\end{equation}
where we have collected all frequency and time related terms under the same source-time function 
\begin{equation}\label{eq:source_time}
\bar{g}_k(t) = \int_{-\infty}^{t} (\iu \nu_k)^{-1}  \mathrm{e}^{\iu \nu_k (t-t')} \mathrm{e}^{i \omega_g t'} dt'.
\end{equation}
 Next, we insert equation (\ref{eq:eigenfunctions}) in (\ref{eq:disp_tidal}), and we isolate the radial and the angular integrals to arrive at 
 \begin{eqnarray}
 \begin{aligned}
     \mathbf{s}_k (\mathbf{r},t) =  - \frac{1}{2} h_0 \omega_g^2 \mathbf{s}_k (\mathbf{r}) \bar{g}_k(t) \  \bm{\epsilon} & :  \left[ \int_{r'} r'^{3} U_k(r') dr' \int_{\Omega}  \mathbf{\hat{e}}_r \otimes \mathbf{\hat{e}}_r \mathcal{Y}_{\ell m}(\theta, \phi)  d\Omega  \right. \\ 
     & + \left. \int_{r'} r'^{3} k^{-1} V_k(r') dr' \int_{\Omega} \mathbf{\hat{e}}_r \otimes \nabla_{\ell} \mathcal{Y}_{\ell m}(\theta, \phi) d\Omega \right].     
 \end{aligned}
 \end{eqnarray}
 The later involves integration over tensors of rank 2 which summarizes as 
\begin{equation}\label{eq:i1}
    I_1 = \int_{\Omega}  \mathbf{\hat{e}}_r \otimes \mathbf{\hat{e}}_r \mathcal{Y}_{\ell m}(\theta, \phi)  d\Omega = i_1 \delta_{\ell,0}\delta_{m,0} + i_2 \delta_{\ell,2}\delta_{m,0} + i_3 \delta_{\ell,2}
\end{equation}
and 
\begin{equation}\label{eq:i2}
    I_2 = \int_{\Omega} \mathbf{\hat{e}}_r \otimes \nabla_{\ell} \mathcal{Y}_{\ell m}(\theta, \phi) d\Omega = 3(i_2 \delta_{\ell,2}\delta_{m,0} + i_3 \delta_{\ell,2})
\end{equation}
where $\delta_{i,j}$ is Kronecker delta symbol, and $i_1, i_2, i_3$ are matrix expression that can be found in Appendix \ref{matrix_def}. These are further contracted with the polarisation tensor $\bm{\epsilon}$. Since expression $i_1$ contains an identity matrix, the contraction with $\bm{\epsilon}$ equals to 0, thus the dependency on $l=0$ vanish. The results of the contraction provides the pattern function $f^m(e, \lambda, \nu)$ defined as 
\begin{equation}\label{eq:pattern}
\begin{aligned}
f_m (e, \lambda, \nu) & =  \frac{2}{3} \sqrt{\frac{\pi}{5}}  \delta_{m,0} b_1 \sin^2 e \\
				   & + \delta_{m,2} \frac{C}{2} \left[ 4 b_2 \cos e \cos 2 \lambda + b_1 (3 + \cos 2e ) \sin 2 \lambda \right] \\
				   & + \delta_{m,-2} C \left[ b_1 \cos 2 \lambda (\cos^2 e + 1) - 2 b_2 \cos e \sin 2 \lambda\right] \\
				   & + \delta_{m,1} 2 C \sin e \left[ b_2 \cos \lambda + b_1 \cos e \sin \lambda \right] \\
				   & - \delta_{m,-1} 2 C \sin e \left[ b_2 \sin \lambda - b_1 \cos e \cos \lambda \right],
\end{aligned}
\end{equation}
with $C=2\sqrt{\frac{\pi}{15}}$, $ b_1 = \cos 2 \nu - \sin 2 \nu$, $b_2 = \cos 2 \nu + \sin 2 \nu$, where angle $e$ defines the rotation of the GW in $\mathbf{\hat{e}}_{y} \mathbf{\hat{e}}_{z}$-plane, angle $\lambda$ defines rotation in $\mathbf{\hat{e}}_{x} \mathbf{\hat{e}}_{y}$-plane and angle $\nu$ is the rotation angle about the unit vector $\mathbf{\hat{e}}_k$. The final solution of the displacement is
\begin{equation}\label{eq:tidal_final}
    \mathbf{s}_k (\mathbf{r},t) =  -  \frac{1}{2} h_0 \omega_g^2 \bar{g}_k(t) \mathbf{s}_k (\mathbf{r}) \chi_k  f_m(e, \lambda, \nu)  \delta_{\ell,2},  
\end{equation}
with 
\begin{equation}\label{eq:disp_tidal_chi}
    \chi_k  = \int_{r}  r^3 \rho (r) \left(  U_k(r) +  \frac{3}{\sqrt{6}} V_k(r)\right) dr.
\end{equation}
where $k = \{n,2,m\}$ since after the contraction with the GW polarisation tensor the only remaining angular order is quadruple $\ell=2$ \cite{Ben1983,Boughn1984,Khosroshahi1997,Siegel2010,Maggiore2008}.

The final normal mode solution that describes the response of a solid body due to the GW driving force, and in our case tidal monochromatic force, assumes the summation over all $k$ such as 
\begin{equation}
    \mathbf{s} (\mathbf{r},t) =  -  \frac{1}{2} h_0 \omega_g^2 \sum_n \sum_m \bar{g}_n (t) \mathbf{s}_{n,m} (\mathbf{r}) \chi_n  f_m(e, \lambda, \nu),
\end{equation}
which can be further rewritten as 
\begin{equation}  \label{eq:final_to_use}
      \mathbf{s} (\mathbf{r},t)  =   h(t) R_{P}  \sum_n \frac{\omega_g^2}{\omega_n^2 - \omega_g^2 + i\omega_n \omega_g /Q_n} \sum_m  {\mathbf{s}_{n,m} (\mathbf{r}) \over R_{P}} \chi_n f_m(e, \lambda, \nu)  ,
\end{equation}
or alternatively
\begin{equation}  \label{eq:final_to_usem}
       \mathbf{s} (\mathbf{r},t)  =  h(t) R_{P}   \sum_m f_m(e, \lambda, \nu) \sum_{\alpha} \mathcal{P}_{m}^{\alpha} (\theta,\phi) \sum_n { \mathcal{D}_{n}^{\alpha} (r) \over R_{P} } \frac{\omega_g^2}{\omega_n^2 - \omega_g^2 + i\omega_n \omega_g /Q_n} \chi_n ,
\end{equation}
with $h(t) = -  \frac{1}{2} h_0 e^{i \omega_g t}$, assuming $Q_n >> 1$ and only the sinusoidal part of the source-time function $\bar{g}_n(t)$ (see Appendix \ref{source}), and $m = \{-2:2\}$. From equation (\ref{eq:final_to_use}) and (\ref{eq:eigenfunctions}) we can recognize the non-dimensional transfer function defined as 
\begin{eqnarray}\label{eq:transfer}
    \mathbf{T} (\mathbf{r}, f_g) & = &   \sum_n \frac{f_{g}^2}{f_n^2 - f_g^2 + if_n f_g /Q_n} \sum_m  { \mathbf{s}_{n,m} (\mathbf{r}) \over R_{P}}  \chi_n  f_m(e, \lambda, \nu) \ , \\
    \mathbf{T} (\mathbf{r}, f_g) & = & \sum_{m, \alpha} f_m(e, \lambda, \nu) { \mathcal{P}_{m}^{\alpha} (\theta,\phi})   \sum_n { \mathcal{D}_{n}^{\alpha} (r) \over R_{P} } \frac{f_g^2}{f_n^2 - f_g^2 + if_n f_g /Q_n} \chi_n
\end{eqnarray}
that could be compared with the expression derived in \citeA{belgacem2024coupling}. Finally, this gives the response of the lunar "Weber oscillator" as
\begin{equation} \label{eq:final_with_T}
      \mathbf{s} (\mathbf{r},t) =   R_{P}  \ h(t)  \ \mathbf{T} (\mathbf{r}, f_g) =   R_{P} \ h(t) \sum_{m, \alpha} f_m(e, \lambda, \nu) \ { \mathcal{P}_{m}^{\alpha} (\theta,\phi}) \ T^{\alpha}(r, f_g) .
\end{equation}
where $T^{\alpha}(r, f_g)$ is the component transfer function, which remains the only term within the solution depending on the lunar structure through the eigenfrequencies and the radial functions of the lunar modes $\ell=2$. Further, it is clear that mathematically $n$ should go to infinity, however this is not physically possible. Thus, normal mode approach has to be truncated to some high value of $n_{max}$, where the truncated solution is converging to the actual solution in the mean-squared sense. Eventually, normal modes with the high value of $n$ contributed less to the response, so truncated solution is usually very close approximation to the true response. In the next section we adopt a realistic lunar model and examine the implication of the theoretical solution represented by equation (\ref{eq:final_to_use}).                                                                                 
\subsection{Application to a lunar model}
The radially dependent eigenfunctions $U_k (r)$, $V_k (r)$, $W_k (r)$, the eigenfrequencies $\omega_k$ and the quality factors $Q_k$ which are used to define the Green tensor in equation (\ref{eq:green}) and associated eigenfunctions in equation (\ref{eq:eigenfunctions}) have to be calculate for a particular radial model, where this model contains profiles of the compressional and shear velocities and of the density. We chose as a reference lunar model the one from \citeA{Weber2011}. See \citeA{LOGNONNE201565,garcia2019lunar} for review of other models, all based on the inversion of lunar seismic data recorded by the Apollo Passive Seismic Experiment (PSE) \cite{latham1969passive}. 

\begin{figure}[hbt!] 
\centering
\includegraphics[width=1\columnwidth]{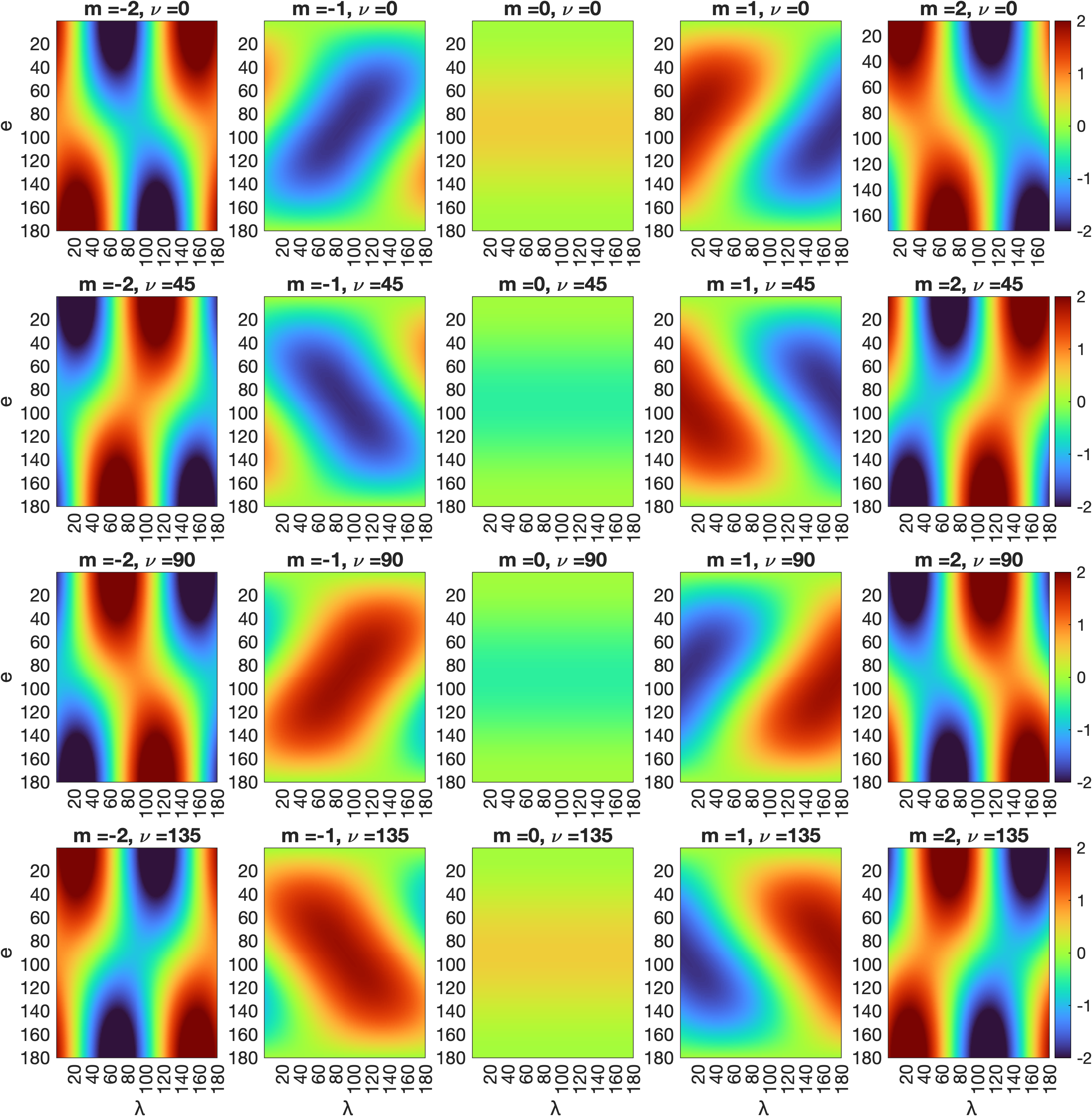}
\caption{Grid search for a pattern function $f_m(e, \lambda, \nu)$ for five azimuthal order $m = \{ -2:2 \}$, and three incident GW angles $\{e, \lambda, \nu \}$ with the tested range $e = \{ 0^{\circ}:180^{\circ}\}$, $\lambda = \{ 0^{\circ}:180^{\circ}\}$,  $\nu = \{0^{\circ}, 45^{\circ}, 90^{\circ}, 135^{\circ}\}$ while keeping the same colorbar for each plot.}
\label{fig:pattern_fun}
\end{figure}

Having normal modes eigenfunction, eigenfrequencies and Q factors calculated for a specific lunar model, we can further explore the implication of the displacement represented by equation (\ref{eq:final_to_use}) by scrutinizing the parameter space spanned by the GW incident angles $\{e, \lambda, \nu \}$, and by the measuring position defined by angles $\{\theta, \phi\}$. The incident angles are the ones defining which azimutal order is going to be excited as shown in Table 1 in \citeA{majstorovic2019earth}. The function $f_m(e, \lambda, \nu)$ that depends on these three angles, we call the pattern function, and it can manifest many combination of angles with maximum, minimum or zero (there is no excitation) values as shown in Figure \ref{fig:pattern_fun}. By performing a gird search analysis over $\{ e, \lambda, \nu \}$ in the range $\{0 - 180^{\circ} \}$ for all azimuthal orders, we obtain that the maximum values per azimuthal orders are for the combination of angles $180^{\circ}$, $4^{\circ}$, $162^{\circ}$ ($m=-2$), $89^{\circ}$, $89^{\circ}$, $113^{\circ}$ ($m=-1$), $89^{\circ}$, $0^{\circ}$, $157^{\circ}$ ($m=0$), $89^{\circ}$, $89^{\circ}$, $22^{\circ}$ ($m=1$), $0^{\circ}$, $22^{\circ}$, $1^{\circ}$ ($m=2$) for $e, \lambda, \nu$, respectfully. In the case of 1D non-rotating lunar model the eigenfrequency for each azimuthal order are the same (i.e. degenerate), since we do not observe the splitting of the quadruple modes. However, in a rotating and non-antisymmetric 3D model for $\ell=2$, the $2\ell+1=5$ singlets will not be degenerated. The percentage of the eigenfrequency splitting have not yet been calculated, but highly likely will not be to same degree as on Earth, due to the lunar slow rotation. Nevertheless, if we also take into account the motion of the Moon within its orbital plane alongside that the eigenfrequencies shall be split in more complex lunar model, we might tackle the ambiguity of the eventual GW sky localization by acknowledging which azimuthal order is excited. 

\begin{figure}[hbt!] 
\centering
\includegraphics[width=1\columnwidth]{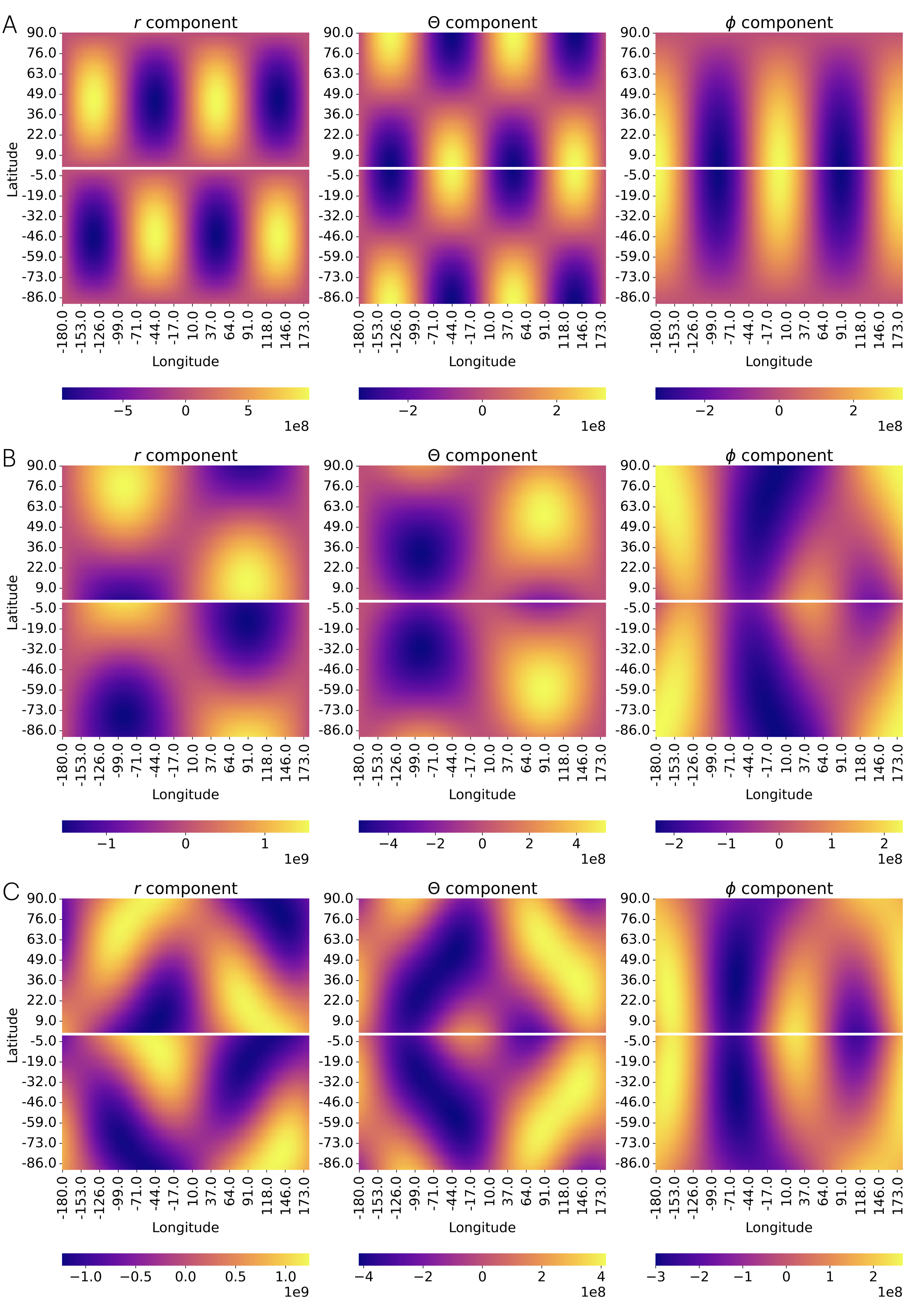}
\caption{Displacement patterns for three components described by equation (\ref{eq:displacement}) by setting the incident GW angles to A) $\{ 1^{\circ}, 0^{\circ}, 22^{\circ} \}$, B) $\{ 90^{\circ}, 0^{\circ}, 0^{\circ} \}$, C) $\{ 45^{\circ}, 0^{\circ}, 0^{\circ} \}$ and by ranging location angles $\{ \theta, \phi \}$ from $\{ -90^{\circ}: 90^{\circ} \}$ and $\{-180^{\circ}:180^{\circ}\}$, respectfully.}
\label{fig:disp_pattern}
\end{figure}

To understand how the displacement depends on the longitude and latitude of the measuring location  when the source is in a given direction we look into the the displacement pattern. We consider the displacement pattern related to the fundamental mode $\omega_{n=0}$ by setting $\omega_g = \omega_{n=0}$ and using specific incident angles, e.g. $\{ e, \lambda, \nu \} = 0^{\circ}, 22^{\circ}, 1^{\circ}$ ($m=2$), and omitting time dependency, therefore from equation (\ref{eq:final_to_use}) we have
\begin{equation} \label{eq:displacement_pattern}
   \mathbf{\tilde{s}} (\theta, \phi)  =  \mathbf{s}_{0,2} (R_{P}, \theta, \phi) \chi_0  f_2 Q_0,
\end{equation}
with $\{\theta, \phi \}$ as latitude and longitude, respectively. In Figure \ref{fig:disp_pattern} we show, for all possible location on the Moon ( i.e. $\{ \theta, \phi \}$ between $\{-90^{\circ}:90^{\circ} \}$ and $\{ -180^{\circ}:180^{\circ}\}$ ), the amplitudes for three different sets of incident angles. This illustrates that for a specific set of incident angles, the pattern displacement per components are distinct. Combining several information that a) different combination of the incident angles excite different azimuthal order, b) different incident angles excite different displacement patterns, c) we expect eigenfrequencies not be degenerated, and d) we anticipate additionally amplitude modulation due to the revolution of the Moon within its orbit, we might be able to associate measurements at one location with one specific GW source. Naturally, this shall be better constrained if we can measure the response on several locations at the lunar surface. 

\begin{figure}[hbt!] 
\centering
\includegraphics[width=0.8\columnwidth]{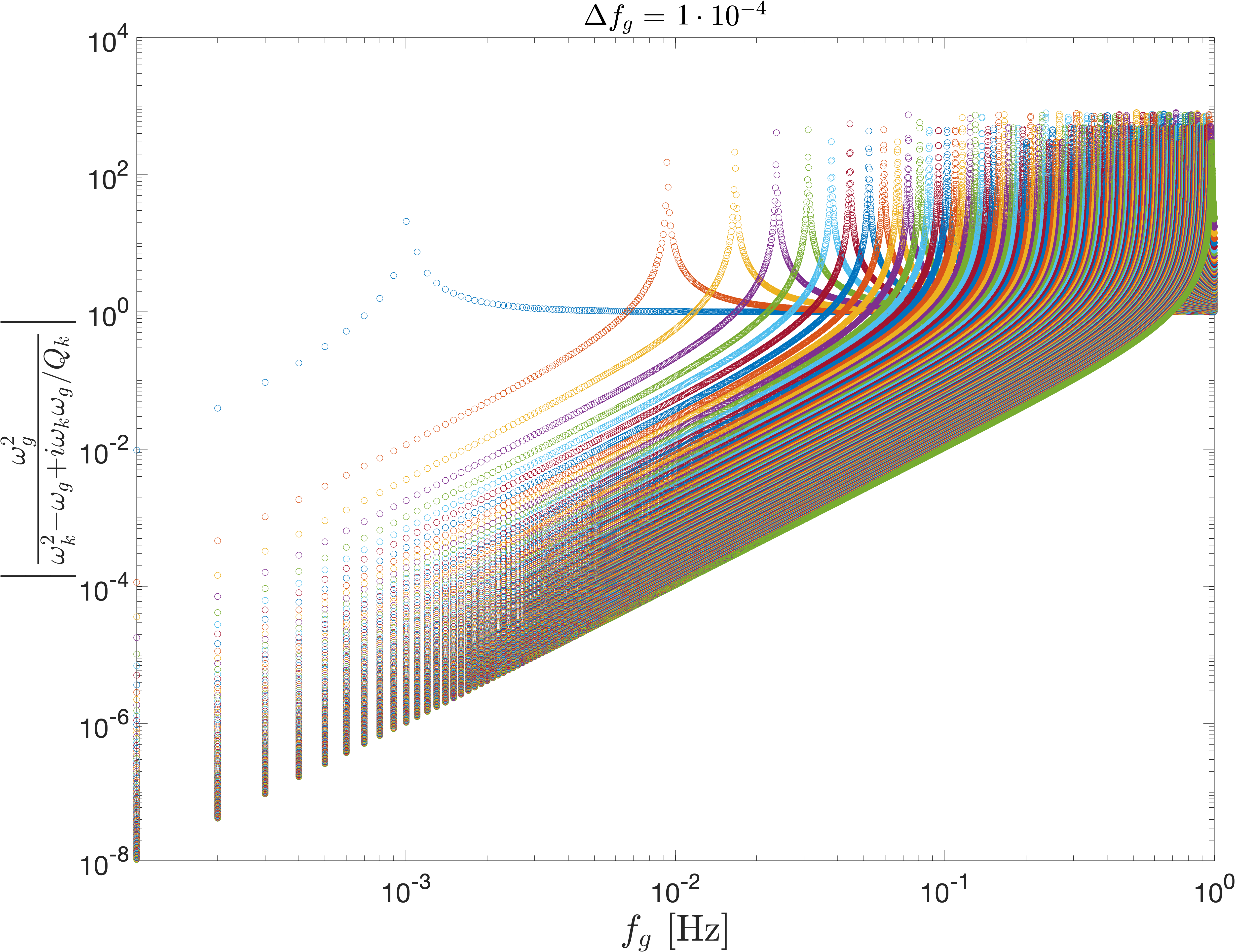}
\caption{Frequency dependent part of the response solution defined by equation (\ref{eq:final_to_use}) obtained by setting the GW frequency resolution $\Delta f_g = 10^{-4}$ Hz and plotting for $n = \{0:10:n_{max} \}$, where $n_{max} \approx 1300$. } 
\label{fig:freq_pattern}
\end{figure}

\begin{figure}[hbt!] 
\centering
\includegraphics[width=1\columnwidth]{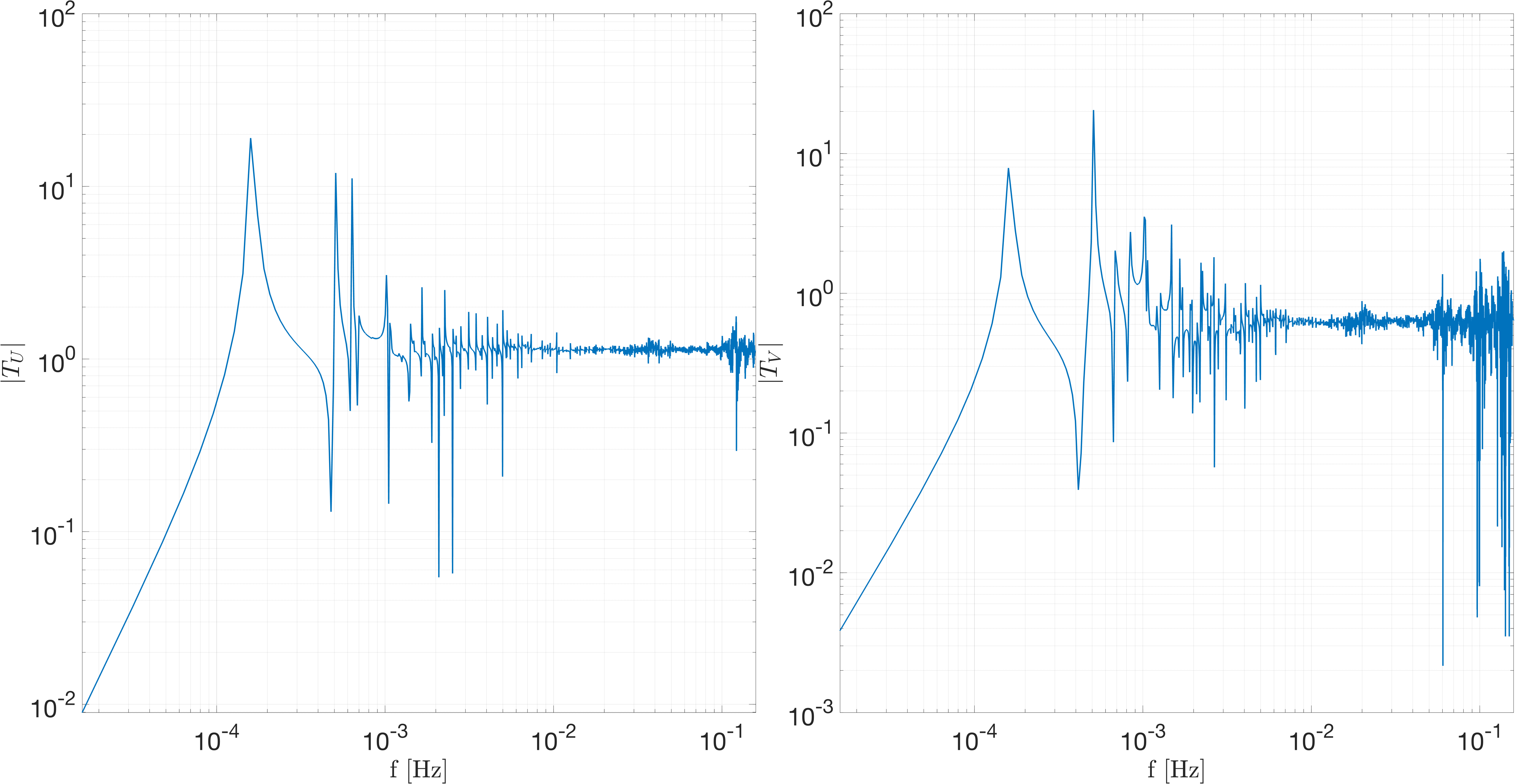}
\caption{Radial transfer functions A) $T_U$ and B) $T_V$ defined by equation (\ref{eq:transfer_radial}) for the frequency range $0-1$ Hz, the frequency resolution $\Delta f_g = 10^{-4}$ Hz and $n_{max} \approx 1300 $.} 
\label{fig:transfer_U_V}
\end{figure}

Next, we can look into frequency dependence of the solution (\ref{eq:final_to_use}) by extracting
\begin{equation} \label{eq:frequency}
    \mathbf{s_f} (\omega_g) =    \sum_n   \frac{\omega_g^2}{\omega_n^2 - \omega_g^2 + i\omega_n \omega_g /Q_n},
\end{equation}
which signify the interplay between the GW frequency and the normal mode eigenfrequencies. We notice that each time we have $\omega_g = \omega_n$ we have resonance as shown in Figure \ref{fig:freq_pattern}. In addition, at the high frequency regime the resonances are very close in frequency space, thus forming a flat level. Eventually, we can plot the transfer function defined with equation (\ref{eq:transfer}) that in this form needs to be calculated for a specific location $\{\theta, \phi\}$ and particular set of incident angles $\{ e, \lambda, \nu \}$ and it can be decomposed in three components. The version of the transfer functions per components is 
\begin{equation}\label{eq:transfer_full}
\begin{aligned}
    \mathbf{T} (r, f_g) = & \; \; \;\; \mathbf{\hat{e}}_r T_U \sum_m \mathcal{Y}_{2,m} f_m(e, \lambda, \nu)   \\
                       & + \mathbf{\hat{e}}_{\theta} T_V \sum_m \frac{1}{\sqrt{6}} \partial_{\theta} \mathcal{Y}_{2,m} (\theta, \phi) f_m(e, \lambda, \nu) \\
                       & + \mathbf{\hat{e}}_{\phi} T_V \sum_m \frac{1}{\sqrt{6}} \frac{1}{\sin \theta} \partial_{\phi} \mathcal{Y}_{2,m} (\theta, \phi) f_m(e, \lambda, \nu)
\end{aligned}
\end{equation}
with the functions that contains sum over the radial order $n$ defined as
\begin{equation}\label{eq:transfer_radial}
\begin{aligned}
        T_U \equiv  \sum_n  \frac{  f_{g}^2  }{f_n^2 - f_g^2 + if_n f_g /Q_n} \frac{\mathcal{U}_n(r)}{R_{P}}  \chi_n , \\
        T_V \equiv \sum_n  \frac{  f_{g}^2  }{f_n^2 - f_g^2 + if_n f_g /Q_n} \frac{\mathcal{V}_n(r)}{R_{P}} \chi_n,     
\end{aligned}
\end{equation}
where the above definitions agree well with \citeA{yan2024toward} to the factor of $1/\sqrt{6}$. To understand the shape of these function it is important to define a relevant frequency band, the frequency resolution $\Delta f_g$ and $n_{max}$. In Figure \ref{fig:transfer_U_V} we plot functions $T_U$ and $T_V$ between 0 and 1 Hz with $\Delta f_g = 10^{-4}$ and $n_{max} \approx 1300 $, which equals to $\approx 0.98$ Hz. We notice that both functions have large peaks in the low frequency regime related to the lowest lunar normal modes, while the response at the high frequency regime becomes flat due to average energy over the high frequency normal modes and effect represented by the equation (\ref{eq:frequency}). The distinguishable peaks are due to the resonance effect, when $2\pi f_g \approx \omega_n$, thus the finer the frequency resolution the narrower are the peaks. However, we also notice the peak associated with the pronounced minimum values between normal mode peaks resonance peaks, which are caused by destructive interference between modes of different radial order $n$. The SI unit of these functions is meter, since functions $U,V$ have units of $kg^{-1/2}$, and $\chi$ has $kg^{1/2}m$. If we want to consider unit-less radial transfer functions $T_U$ and $T_V$, we need to divide them with the normalisation baseline for the meters, which is the radius of the Moon ($\approx 1.737 \cdot 10^6$ m), as it has been done in equations (\ref{eq:transfer_radial}) by introducing $R_P$ parameter. 
Equations (\ref{eq:transfer_radial}) are the most general equations that can be used to calculate the lunar response. To have a response solution with a precise location and incident angles one need to use equation (\ref{eq:transfer_full}). Further, if one wants to calculate the lunar displacement (time evolution) due to incident GW then the appropriate equation is given by relation (\ref{eq:final_to_use}), while in this form it is defined for a monochromatic source. In the case one wish to calculate the response for a GW source other than monochromatic, one needs to redefine equation (\ref{eq:source_time}). 

\begin{figure}[hbt!] 
\centering
\includegraphics[width=1\columnwidth]{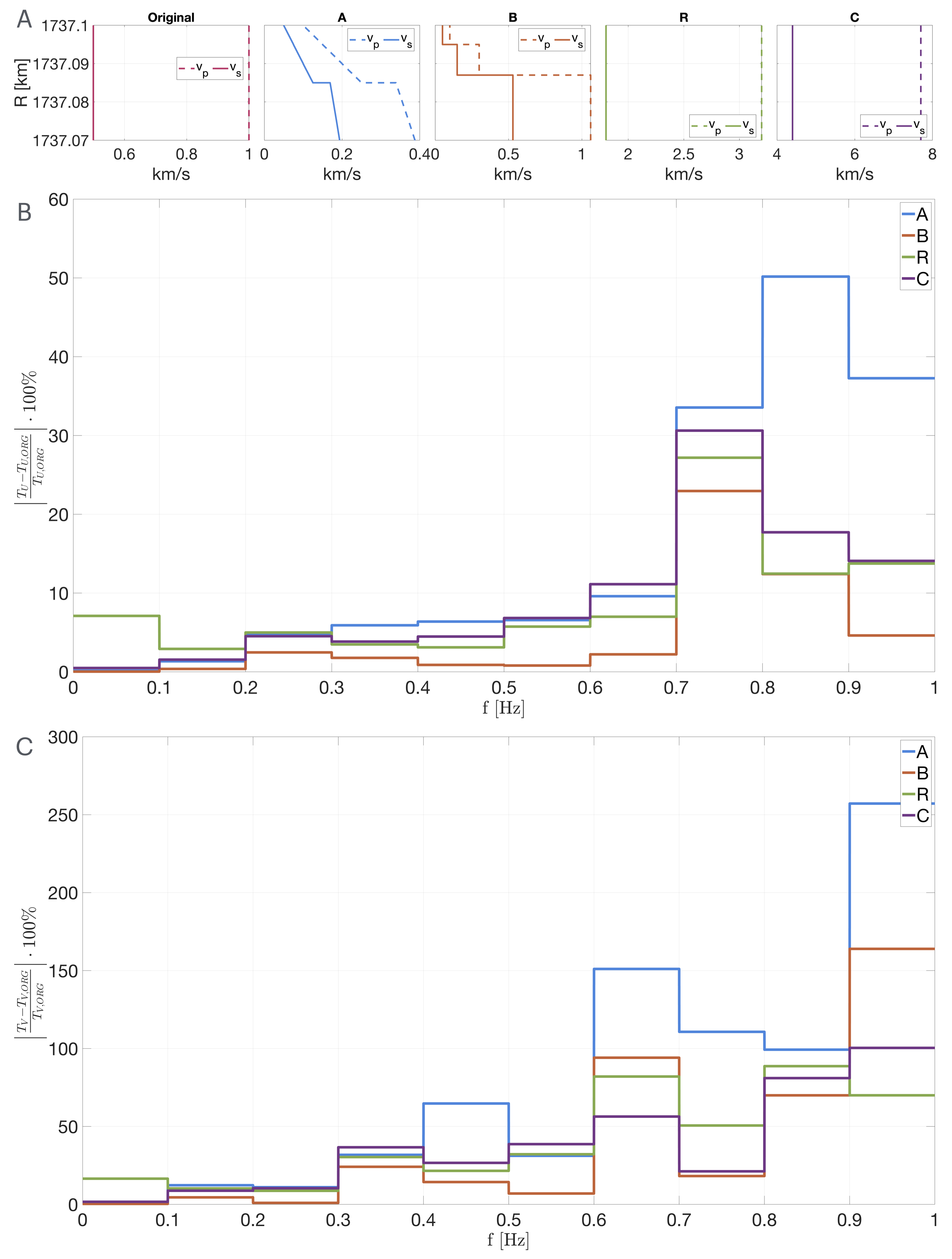}
\caption{A) Shear and compressional velocity profiles first few meters for several models. B) Relative error estimates for radial transfer functions $T_U$ calculated for the models from A) where $T_{U, ORG}$ is calculated for the lunar model from \citeA{Weber2011}. C) Same as B) just for radial function $T_V$. All radial tranfer functions are calculated in the frequency range $0-1$ Hz, with the frequency resolution $\Delta f_g = 10^{-4}$ Hz and $n_{max} \approx 1300 $.} 
\label{fig:models}
\end{figure}

One must take into account that the surface layer of the Moon is much more scattered than the one of the Earth's \cite{troitskii1962nature, toksoz1974structure, shkuratov2001regolith, lucey2006understanding}. To account for the scattering effect and to study how the regolith structure might affect the GW response, next we modify the first few meters of the original lunar model and introduce four new models. This implies chaining the compressional and shear velocity profiles. The labels associated with the models are: A for a model with the linear increase of the velocities, B for a model with step-like features, R for a model without regolith structure (value of the velocities are the same in the first 15 km), C for a model without crust (meaning that the first discontinuity we have below 1000 km) as shown in Figure \ref{fig:models} A. In Figure \ref{fig:models} B and C we show the results for radial transfer functions $T_U$ and $T_V$, respectively. However, for better visual clarity, instead of plotting the actual radial transfer functions we calculate the relative error with respect to the original transfer function associated with the original model in Figure \ref{fig:models} A. Further, the relative error is averaged in ten bins defined as $\{(i-1)\cdot 0.1, i \cdot 0.1 \}$ for $i =1, ..., 10$. We can notice how both functions are affected by modifying a thin layer at the lunar surface, where function $T_V$ is more affected than $T_U$ indicated by larger relative errors for all models. We can also notice that overall relative error is much larger in the high frequency regime, for $T_U$ function the boundary is defined at 0.7 Hz, while for $T_V$ function around 0.6 Hz. Next, both functions show almost no change in the lowest frequency regime, expect for model R ($\approx 20 \%$). Also, for both functions the highest total relative error is associated with the model A, while for function $T_U$ next is model C, model R, and finally model B, and for function $T_V$ the second highest total relative error is associated with model R, then model B and model C. The structure of the thin layer of model A, probably produces surface trapped modes which are dominating in the high frequency regime. This implies that modeling GW response in the high frequency domain requires finer scale modeling of the surface lunar velocity layer. There is a high possibility that with the data from the future lunar missions, specially Farside Seismic Suite mission \cite{panning2021farside} which shall deploy a seismometer at the South pole, we can put constrains on the surface velocity profiles. Consequently, this means that we can obtain more consistent analytical lunar GW response in the high frequency regime.   

Next, we discuss the existing lunar response models from the literature and 
present some perspectives on the GWs detectability. 

\clearpage

\section{Discussion}
\subsection{On the existing lunar response models}
As mentioned earlier several papers has been addressing the solution of the lunar response due to the GWs \cite{harms2021lunar, branchesi2023lunar, li2023detecting, kachelriess2023lunar, cozzumbo2023opportunities, yan2024toward, bi2024response, ajith2024lunar}. Earlier, there have not been enough clarity about which solution should be considered in the context of the coordinate system one should adopt based on the measuring principle one imposes \cite{harms2022seismic}. In general, it is important to understand the correct way the elastic media is coupled with the GWs, which eventually redefines the equation of motion and it's solution \cite{ashby1975gravitational, belgacem2024coupling}. From the current status of the lunar response papers, we can group them by the solution they adopt, one based on the principle introduced in paper by \citeA{Dyson1969}, where GW force is defined as a gradient of the shear modulus, or one where GW force is Newtonian tidal one. The papers that discuss Dyson response solution are \citeA{branchesi2023lunar, kachelriess2023lunar, bi2024response}, those that discuss tidal one are \citeA{harms2021lunar, li2023detecting, cozzumbo2023opportunities}, and finally paper that discusses both responses is \citeA{yan2024toward}.

We shall notice that all provided solutions uniformly agree that the only quadruple ($l=2$) normal modes are going to be excited by passing GWs. The overall analytical solutions bear some differences, mainly due to spherical harmonics or eigenfunction normalisations and the equation of motions one is solving for. In the normal mode theory eigenfrequencies and radial eigenfunctions are usually solved for spherically symmetric non-rotating model with isotropic elastic tensor by the numerical solver such as is MINEOS \cite{mineos-v1.0.2}. We essentially are solving the equations of motion where we do account for the gravitational potential and the gravity field. Next, the choice of the lunar model, the radial velocity and density profiles, is also highly important. Additionally, the response solution depends on the number of layers in the model and the radial order $n$, which was rightfully recognized in other studies as well. Eventually, more realistic and layered lunar models with higher radial order $n$ define more detailed structure in our transfer functions compared to some previous works.

\citeA{harms2021lunar} discusses the lunar response due to the Newtonian tidal force based on the frequency dependency introduced in their transfer function. They do not implement a particular lunar model to obtain the final solution, but the paper discusses how the lowest normal modes should have Q values around 200, and that the effective baseline, e.g. the size, should be around $0.6 R_{Moon}$. They motivated truncating their solution to $n=22$, indicating that in the high frequency regime we have reduced response due to the partial cancellation of the baseline $L_n$ in the sum. Their solutions is flat in the full frequency regime, which agrees with ours, beside the point that not having a lunar model they lack the structure in the normal modes spectrum in the low and high frequency regimes. 

\citeA{branchesi2023lunar} make use of the Dyson definition of the GW force, and followed the formalism from paper \citeA{majstorovic2019earth}. They used the same lunar model as in this paper. We observe the structure within the GW response by utilizing the realistic lunar model, however these solutions falls off as $1/f$ as Dyson predicted, and thus is not directly comparable with our transfer function. 
 
\citeA{kachelriess2023lunar} implemented their own solver for eigenfrequencies and eigenfunctions values for three different lunar models from \citeA{garcia2019lunar}. They reported that since the largest differences between the models are related to the core, the variations between eigenfrequencies are increasing with radial order $n$, achieving $20\%$ difference for $n = 4$. They also adopted a constant Q for all eigenfrequecies of 3300, which should be too high for a realistic observable Q. The overall response function in their study cannot be directly compared to ours, since they consider the Dyson principle, which is reflected within their source time function. They also rightfully addressed the question of the displacement angular patterns related to three components.  

\citeA{li2023detecting} do consider a Newtonian tidal forcing, however we do not agree with the content of the transfer function (their relation [14] shall be compared with equations (\ref{eq:transfer_radial})), where in our case we have an additional radial eigenfunction in the numerator. They solve the equation of motion without gravity and gravitational potential, also deploying different eigenfunction normalisation (see [a23]). Their response function is derived without considering radial velocity profile, but selecting a value of 3000 km/s for a local sound velocity, Poisson's ratio of 1/3, $Q = 300$ for all modes, and truncating $n=30$, motivated by the fact that high-order normal modes are severely suppressed. However, as showed in our study the high frequency domain, around 1 Hz, is not suppressed and it predominately depends on the first few meters of the velocity profile and truncation is important if we want to have well founded response function in the high frequency regime. Since they do not deploy the radial model, their GW response function is rather smooth without complexity as presented in our Figure \ref{fig:transfer_U_V}. 

\citeA{yan2024toward} study follows mathematical formalism from \citeA{majstorovic2019earth}, however using different eigenfunction normalisation. Nevertheless, their radial transfer function matches ours (compare their equations [23], [24] with relation (\ref{eq:transfer_radial})) to the factor of $1/\sqrt{6}$. To compute the GW transfer function they use the lunar model compiled from several published lunar models, for $Q_{core}$ they settled for 1000, and they truncated radial order to $n=400$. We can observe that our transfer functions match in the low frequency regime, however somewhere around $10^{-2}$ Hz solutions start to mismatch. While our solution is flat, theirs falls off in the high frequency regime. They claim that their high frequency regime is affected by the artificial truncation of the normal modes to $n=400$ which should translate to 0.32 Hz. Even if we truncated our response function to $n=400$, we still keep flat frequency response. We believe that the shape of the transfer function is governed by the relation (\ref{eq:frequency}), which clearly has a flat response in the whole frequency regime. Nevertheless, we notice that their response function also reflects the complex structure of the realistic lunar model. 

\citeA{cozzumbo2023opportunities} most likely follows the same principles as in \citeA{harms2021lunar} due to the same shape of the function and logic discuss as in the referred paper. 

\citeA{bi2024response} considers Dyson definition of the force and eventually studied the effect of the horizontally layered half space. The study shows how important is to have radially heterogeneous model, which brings up the complex structure in the GW response spectrum. However, their simulation predicts lower response compared to the normal-mode simulation. Which might emphasise the importance of having a finite size model. They emphasised the importance of constraining the surface layer of the shear-wave velocity, since it has a strong impact on the lunar GW response, the slower the velocity the larger the response is. This agrees with our results in Figure \ref{fig:models}, since models labelled as "original" and "A" have the lowest values of the shear velocity, but the largest variations in the high frequency regime. They also showed that high Q values, should be responsible for GW response amplifications. 

All the studies that provide their response functions in SI units, seem to agree on the order of the amplitude around the lowest normal mode which translates to the frequency around $10^{-3}$ Hz. The order of the amplitude ranges between $10^7-10^8$ m, and can be interpreted as an amplification factor, since this factor should be multiply with the scalar of the metric perturbation $h_0$ to obtain the value that should be measured by the proposed instruments. In the high frequency regime we noticed that the order of the amplitude changes depending on the proposed force model, Dyson or tidal. We either expect for this response to fall of as $1/f$ as predicted by Dyson force term or to be flat for the tidal one.  

While the order of the amplitude during the whole frequency range might be determined by the proposed models, it is important to emphasise how the detailed structure still remains unknown. This is directly related to the lunar model, the radially and laterally heterogeneous profiles, which still needs to be determined directly from the observations. There are many prospects why this might be possible in the future with the missions such as Lunar Geophysical Network \cite{haviland2022lunar}, Farside Seismic Suite \cite{panning2021farside}, Chang’e 7 \cite{wang2024scientific}, where one of the scientific payloads are seismometers. If we would be able to refine the 3D structure of the lunar models, this implies that we can improve our normal-mode simulations by introducing coupling and splitting of the modes.

\subsection{On the detectabilty}
In this paper we focus solely on the model that describes the interaction between GWs and the elastic bodies, such as is the Moon in rough approximation. From the previous chapter, we notice that this model is just an approximation of the more compound interaction, where further improvements in geophysical and relativistic sense are still to be made \cite{Dahlen1998, belgacem2024coupling}. Considering spherically symmetric non-rotating planet and using non-relativistic theory of elasticity in the TT frame, we arrive at the transfer function described with equation (\ref{eq:transfer_full}). This transfer function gives the displacement field in meters relative to the center of mass of the Moon and induced by a GW of strain $h=1$, which divided by the baseline that is the lunar radius, becomes unit-less. Having the function in this shape is adapted to the strainmeter measurements \cite{ashby1975gravitational}. In the case where detection is done with seismometers, one needs to use equation of motion (\ref{eq:motion_TT_non_relat}) naturally defined in the TT coordinates, which directly provides the evolution of the observable for an ideal seismometer, and derive the appropriate transfer function for it. However, to actually measure the GW induced displacement field we need to understand how sensitive the instruments are and how they couple with the environment. For this purpose, an analysis of all noise sources coming from the environment and the instrument itself coupling with the measurement needs to be investigated in order to get an instrumental sensitivity curve \cite{cozzumbo2023opportunities}. Having the transfer function $\mathbf{T}$ (in m) of the lunar response with the instrument noise spectrum density $\mathbf{n_f}$ (in $\mathrm{m/\sqrt{Hz}}$), we can obtain the general shape of the characteristic noise strain $\mathbf{h_{n}}$ \cite{Moore2015} for our detector as 
\begin{equation}
    h_{n}=\frac{n_f\sqrt{f}}{T}. 
\end{equation}
where the transfer function $\mathbf{T}$ can be either for radial or horizontal direction as shown in equation (\ref{eq:transfer_radial}).
Next important ingredient on the topic of GW detectability is the lunar environment. It is well known that Moon is seismically active body, which can host shallow moonquakes, meteorid impacts and deep moonquakes \cite{nunn2023moon}. However, as discussed before, it is known that its seismicity upper limit in the deci-Hertz band is 3 orders of magnitude lower than the Earth one \cite{https://doi.org/10.1029/2008JE003294}. During the seismically active period the detection of gravitationally induced lunar response would not be possible. Therefore, one strategy might be to search for and analyse only the aseismic time periods. Moreover, during these phases we should be naturally limited with the seismic hum, everlasting lunar excitation govern by deep moonquakes \cite{https://doi.org/10.1029/2008JE003294, lognonne2015impact}. One possible difficulty on the Moon should be the extreme temperature variations \cite{williams2019seasonal}, which probably will have to be compensated with proper instrument shielding, and careful modeling of the thermally induced elasticity of the ground, or selecting landing site where these variations might be the minimal. 
Finally, we should be aware of the types of sources we are able to measure, with the future lunar GW detectors, based on their frequency content and amplitude. Since the substantial lunar amplification due to GWs occurs between the mHz and deciHz bands, the future lunar detectors overlap in great extent with the sources commonly associated with the LISA mission  \cite{colpi2024lisadefinitionstudyreport}. The most intense sources expected in this frequency band are Super Massive Black Holes Binaries (SMBHB), therefore those source might be most probably sources to capture.

\section{Conclusion} 
In this paper, we revisit the theory behind the gravitational waves (GWs) induced response of a spherically symmetric, non-rotating body, in this particular case the Moon. After reviewing how GWs appear in the equations of motion using non-relativistic elasticity theory, both in the Transverse-Traceless frame and the proper detector frame, we discuss the link between these approaches and how they define their solutions. The derived analytical solution describes how the Moon shall respond to passing GWs in terms of the normal modes approximation. The response model is naturally parameterised into several components, which can be further studied to optimise the eventual GW detection. In essence, these components are the incident angles of the GWs, the measuring location, the source time function. Exploring the parameter space of the interaction between GWs and the Moon we examine the transfer function. We show that this transfer function has a relevant frequency band defined by the normal-modes of the Moon. Moreover, the transfer function depends on the lunar model, defined by the compressional, shear velocities and density radial profiles. Altering very shallow structure of the original lunar model, we show how this eventually alters the high frequency regime of the transfer function between 0.1 and 1 Hz. This directly has implications for the possible detection, and requires constraining the shallow regolith models with geophysical methods. Furthermore, to evaluate the relevance of a possible gravitational wave detection on the Moon, we need to further improve the analytical solution by including the 3D lunar models, but also derive a full instrumental noise budget to obtain a sensitivity curve in the context of the lunar environment.  

\acknowledgments
Authors would like to thank the French Space Agency, CNES, for supporting this research in the frame of the French contribution to FSS (Farside Seismic Suite) as well as IDEX Paris Cité (ANR-18-IDEX-0001). 

\clearpage
\section*{Appendix}
\setcounter{equation}{0}
\renewcommand{\thesubsection}{\Alph{subsection}}
\renewcommand{\theequation}{\thesubsection.\arabic{equation}}

\subsection{Christoffel symbol in TT gauge}
\noindent
 Christoffel symbol $\Gamma_{\nu \rho}^\mu$  can be written as
 \begin{equation}\label{eq:christoffel}
     \Gamma_{\nu \rho}^\mu=\frac{1}{2} g^{\mu \sigma}\left(\frac{\partial g_{\rho \sigma}}{\partial x^{\nu}}+\frac{\partial g_{\nu \sigma}}{\partial x^{\rho}}-\frac{\partial g_{\nu \rho}}{\partial x^{\sigma}}\right).
 \end{equation}
 With a metric $g_{\mu \nu}=\eta_{\mu \nu}+\boldsymbol{h}_{\mu \nu}$, the Christoffel symbol can
 restricted to first order in $h_{\mu \nu}$ as 
\begin{equation}
    \Gamma_{\nu \rho}^\mu=\frac{1}{2} \eta^{\mu \sigma}\left(\partial_\nu h_{\rho \sigma}+\partial_\rho h_{\nu \sigma}-\partial_\sigma h_{\nu \rho}\right).
\end{equation}
Because both $h_{00}$ and $h_{0 i}$ are set to zero in TT gauge, one can find
\begin{equation}\nonumber
    \Gamma_{00}^i=\frac{1}{2}\left(2 \partial_0 h_{0 i}-\partial_i h_{00}\right) = 0
\end{equation}
and the only non-vanishing terms
\begin{equation}\nonumber
    \Gamma_{0 j}^i=\frac{1}{2} \partial_0 h_{i j}. 
\end{equation}

\setcounter{equation}{0}
\subsection{TT to LL transformation}\label{annexe:TT_to_LL}
We want to pass in Fermi coordinates to establish a local inertial frame at Moon center of mass and describe the gravitational wave coupling as a Newtonian force in the detector frame. For this purpose, we use the same method used in \citeA{ashby1975gravitational} which consist on changing coordinates in two steps.
One can define new variables $\bar{x}$ with a quantity $\epsilon(\bar{x})$ of same magnitude of $h$. Then, for the rest of the computation we only consider the linear terms in $\epsilon(\bar{x})$. Thus we have
\begin{equation}\label{first_coordinate_change}
    \begin{aligned}
& x_0=\bar{x}_0, \\
& x_3=\bar{x}_3, \\
& x_a=\bar{x}_a+\epsilon_a(\bar{x})
\end{aligned}
\end{equation}
with $x_\mu=\eta_{\mu \nu} x^\nu$.
\\
Writing again the metric in TT with the new coordinate gives us
\begin{equation}
\begin{aligned}
d s^2= & \eta_{\mu \nu} d \bar{x}^\mu d \bar{x}^\nu+2 \epsilon_{a, 0} d \bar{x}^a d \bar{x}^0+\left(\epsilon_{a, b}+\epsilon_{b, a}\right) d \bar{x}^a d \bar{x}^b \\
& +h_{a b} d \bar{x}^a d \bar{x}^b+2 \epsilon_{a, 3} d \bar{x}^a d \bar{x}^3.
\end{aligned}
\end{equation}
In order to get rid of the $d \bar{x}^a d \bar{x}^b$ terms we choose $\epsilon_{a}$ that satisfies expression
\begin{equation}\label{eq:h_r_s}
    h_{a b}=-\left(\epsilon_{a, b}+\epsilon_{a, b}\right).
\end{equation}
\textcolor{black}{\noindent
To resolve this partial differential equation, we start by looking for a homogeneous solution $\varepsilon^{(h)}$ with  $h_{a b}= 0$ as
\begin{equation}
    \varepsilon^{(h)}_a = H(x^0,x^3)
\end{equation}
with $H$ a function depending only on $x^0$ and $x^3$.
\\
A particular solution $\epsilon^{(h)}$ can be chosen as a first-order polynomial as
\begin{equation}
    \epsilon^{(p)}_a = P_{al}(x^0,x^3)x^l,
\end{equation}
with $P_{al}=P_{la}$ a symmetrical tensor depending only on $x^0$ and $x^3$. \\
The general solution $\epsilon_{a} = \epsilon^{(p)}_a + \epsilon^{(h)}_a$ can be injected in \ref{eq:h_r_s} to obtain
\begin{equation}
    \begin{aligned}
h_{a b} & =-\left(\frac{\partial \epsilon_a}{\partial x^b}+\frac{\partial \epsilon_b}{\partial x^a}\right) \\
& =-\left(P_{a l}\left(x^0,x^3\right) \delta_b^l+P_{b l}\left(x^0, x^3\right) \delta_a^l\right) \\
& =-2 P_{a b}\left(x^0,x^3\right)\\
&\Rightarrow P_{a b} = -\frac{1}{2}h_{a b}.
\end{aligned}
\end{equation}
In order to minimize the number of new terms, we choose to have $\epsilon_{a} = 0$ at the coordinate frame origin ($x^\mu=0$) and obtain a general solution for \ref{eq:h_r_s} with $H(x^0,x^3)=0$ as
\begin{equation}
    \epsilon_a=-\frac{1}{2} h_{a b} x^b.
\end{equation}}
\\ 
This first transformation considerably simplified the notations. The resulting metric, if we omit bars on new coordinates, takes the form of
\begin{equation}\label{metric_after_first_trans}
    d s^2=\eta_{\mu \nu} d x^\mu d x^\nu+2 \epsilon_{a, 0} d x^a d x^0+2 \epsilon_{a, 3} d x^a d x^3.
\end{equation}
\\
The second step of gauge transformation consists on defining new coordinates as 
\begin{equation}
    \begin{aligned}
& x_0=\bar{x}_0+\epsilon_0(\bar{x}), \\
& x_3=\bar{x}_3+\epsilon_3(\bar{x}), \\
& x_a=\bar{x}_a.
    \end{aligned}
\end{equation}
\\
We write equation \ref{metric_after_first_trans} with new coordinates to obtain
\begin{equation}
    \begin{aligned}
        ds^2 = &\eta_{\mu\nu}d\bar{x}^{\mu}d\bar{x}^{\nu} + \left( 2\epsilon_{a,0}+ 2\epsilon_{0,a}\right)d\bar{x}^{a}d\bar{x}^{0} + \left( 2\epsilon_{a,3}+ 2\epsilon_{3,a}\right)d\bar{x}^{a}d\bar{x}^{3} \\
        &+ \left( 2\epsilon_{0,3}+ 2\epsilon_{3,0}\right)d\bar{x}^{3}d\bar{x}^{0} + 2\epsilon_{0,0}d\bar{x}^{0}d\bar{x}^{0} +  2\epsilon_{3,3}d\bar{x}^{3}d\bar{x}^{3}.
    \end{aligned}
\end{equation}
As if it was in the first step, we choose $\epsilon$ to simplify this last equation and to get rid of $d \bar{x}^0 d \bar{x}^a$ and $d \bar{x}^a d \bar{x}^3$ terms. For this, the condition is written as
\begin{equation}\label{eq:step2_eps}
    \epsilon_{0, a}=-\epsilon_{a, 0}, \quad \epsilon_{3, a}=-\epsilon_{a, 3}.
\end{equation}
\noindent
To resolve $\epsilon_{0, a}=-\epsilon_{a, 0}$, we can assume that our solution is a second order polynomial as
\begin{equation}\label{eq:step_2_choice}
    \epsilon_0=H_{kl} x^k x^l
\end{equation}
with $H_{kl}=H_{lk}$ a symmetrical tensor depending only on $x^0$ and $x^3$. Injecting this solution in the left part of equation \ref{eq:step2_eps}, we obtain
\begin{equation}
\begin{aligned}
\frac{\partial \epsilon_0}{\partial x^a} & =H_{k l} \delta_a^k x^l+H_{k l} x^k \delta_a^l \\
& =H_{a l} x^l+H_{k a} x^k. \\
\end{aligned}
\end{equation}
Here, $l$ and $k$ are dummy indices so one can write
\begin{equation}
    \frac{\partial \epsilon_0}{\partial x^a}  =2 H_{a l} x^l.
\end{equation}
\\
Using this last result and $\epsilon_a=-\frac{1}{2} h_{a l} x^l$, we can solve conditions \ref{eq:step2_eps} as
\begin{equation}
    \begin{aligned}
 2 H_{a l} x^l&=\frac{1}{2} \frac{\partial}{\partial x^0}\left(h_{a l} x^b\right)=\frac{1}{2}  \left(\frac{\partial h_{a l}}{\partial x_0} x^l+ \underbrace{h_{a l} \delta_0^l}_0\right) \\
\Rightarrow H_{al} x^l &=\frac{1}{4} h_{al, 0} x^l \\
\Rightarrow H_{al}&=\frac{1}{4} h_{al, 0}. \\
    \end{aligned}
\end{equation}
One can replace this last expression in equation \ref{eq:step_2_choice} to get expressions such as 
\begin{equation}
    \epsilon_0=\frac{1}{4} h_{a b, 0}x^ax^b,
\end{equation}
and 
\begin{equation}
    \epsilon_3=\frac{1}{4} h_{a b, 3}x^ax^b.
\end{equation}
Finally, we get a simplified form of the metric depending on the second derivatives of $h_{a b}$ as
\begin{equation}
    \begin{aligned}
d s^2= & -\left(d \bar{x}^0\right)^2\left(1-\frac{1}{2} h_{a b, 00} x^a x^b\right)+\left(d \bar{x}^3\right)^2\left(1+\frac{1}{2} h_{a b, 33} x^a x^b\right) \\
& +2\left(\frac{1}{2} h_{a b, 03} x^a x^b d \bar{x}^0 d \bar{x}^3\right)+\left(d \bar{x}^1\right)^2+\left(d \bar{x}^2\right)^2.
    \end{aligned}
\end{equation}

\setcounter{equation}{0}
\subsection{The real spherical harmonics}\label{harmonics}
We follow the definition of the real spherical harmonics given in book \citeA{Dahlen1998} as
\begin{equation}\label{eq:th4}
\mathcal{Y}_{lm}(\theta, \phi) = \left\{\begin{array}{l} \sqrt{2} X_{l|m|} (\theta) \cos m \phi \;\; \text{if} \;\; -l \le m < 0 \\ X_{l0} (\theta) \;\; \text{if} \;\;  m = 0  \\ \sqrt{2} X_{lm} (\theta) \sin m \phi \;\; \text{if} \;\; 0 < m \le l  \end{array}\right.
\end{equation}
with 
\begin{equation}\label{eq:th5}
X_{lm} = (-1)^{m} \sqrt{\frac{2l+1}{4 \pi}} \sqrt{\frac{(l-m)!}{(l+m)!}} P_{lm} (\cos \theta).
\end{equation}

\setcounter{equation}{0}
\subsection{Source time function}\label{source}
To solve source-time function for the monochromatic source defined as
\begin{equation}\label{eq:source1}
\bar{g}_k(t) = \int_{-\infty}^{t} (\iu \nu_k)^{-1}  \mathrm{e}^{\iu \nu_k (t-t')} \mathrm{e}^{i \omega_g t'} dt',
\end{equation}
we use the convolution theorem where we separate two functions $f_1(t)$ and $f_2(t)$ as 
\begin{equation}\label{eq:c2}
\begin{aligned}
    f_1 (t) & = \mathrm{e}^{i \omega_g t'}, \\
    f_2 (t) & = (\iu \nu_k)^{-1}  \mathrm{e}^{\iu \nu_k (t-t')}  \\
            & \approx \frac{1}{\omega_k^2} \frac{1}{1+ \frac{1}{4Q_k^2}} (\omega_k \sin (\omega_k t) - \gamma_k \cos(\omega_k t)) \mathrm{e}^{-\gamma_k t }
\end{aligned}
\end{equation}
where for function $f_2(t)$ we took only the real part and we used the definition $\nu_k = \omega_k + \iu \gamma_k$. 
The Fourier transform of these functions are 
\begin{equation}
\begin{aligned}
    \mathcal{F} \{ f_1 (t) \} & = \delta (\omega - \omega_g), \\
    \mathcal{F} \{ f_2 (t) \} & = \frac{1}{\omega_k^2} \frac{1}{1+ \frac{1}{4Q_k^2}}  \frac{\omega_k^2 - \gamma_k^2 + \iu \gamma_k \omega }{\gamma_k^2 + 2\iu\omega \gamma_k - \omega^2 + \omega_k^2 }. \\
\end{aligned}
\end{equation}
Therefore, solving equation (\ref{eq:source1}) with convolution theorem yields 
\begin{equation}
\begin{aligned}
    \bar{g}_k(t) & = \frac{1}{2 \pi} \int_{-\infty}^{+\infty} \mathcal{F} \{ f_1 (t) \} \cdot \mathcal{F} \{ f_2 (t) \} \mathrm{e}^{\iu \omega t} d \omega  ,\\
     &= \frac{1}{2 \pi} \frac{1}{\omega_k^2} \frac{1}{1+ \frac{1}{4Q_k^2}}  \frac{\omega_k^2 - \gamma_k^2 + \iu \gamma_k \omega_g }{\gamma_k^2 + 2\iu\omega_g \gamma_k - \omega_g^2 + \omega_k^2 }. 
\end{aligned}
\end{equation}
Next, if we consider an approximation where $Q_k \gg $ we arrive at 
\begin{equation}
    \bar{g}_k(t) \approx \frac{1}{2 \pi} \frac{1}{\omega_k^2}  \frac{\omega_k^2 + \iu \gamma_k \omega_g }{\omega_k^2 + 2\iu\omega \gamma_k - \omega_g^2}.
\end{equation}
This expression differs a bit from \citeA{yan2024toward}, thus if we only consider sinusoidal function in expression (\ref{eq:c2}) we get 
\begin{equation}
    \bar{g}_k(t) \approx \frac{1}{2 \pi}  \frac{1 }{\omega_k^2 + 2\iu\omega_g \gamma_k - \omega_g^2},
\end{equation}
the expression for the source-time function which is eventually used in this study. 

\setcounter{equation}{0}
\subsection{Matrix definition}\label{matrix_def}

\begin{equation}
    i_1 = \frac{2 \sqrt{\pi}}{3}  \begin{bmatrix}
1 & 0 & 0 \\
0 & 1 & 0 \\
0 & 0 & 1
\end{bmatrix}
\end{equation}

\begin{equation}
    i_2 = \frac{2}{3} \sqrt{\frac{\pi}{5}} \begin{bmatrix}
-1 & 0 & 0 \\
0 & -1 & 0 \\
0 & 0 & 2
\end{bmatrix}
\end{equation}

\begin{equation}
    i_3 = 2\sqrt{\frac{\pi}{15}} \begin{bmatrix}
\delta_{m,-2} & \delta_{m,2} 	& -\delta_{m,-1} \\
\delta_{m,2}  & - \delta_{m,-2} & -\delta_{m,1}  \\
-\delta_{m,-1} & -\delta_{m,1} 	& 0
\end{bmatrix}
\end{equation}

\bibliography{main}

\end{document}